# Satellite gravity constraints on inner core viscosity and LLVPs density anomalies

Yachong An[1], Hao Ding[1*], Fred D. Richards[2], Weiping Jiang[3], Jiancheng Li[4], Wenbin Shen[4]

[1]*School of Earth and Space Science and Technology, Wuhan University, 430072, Wuhan, China*

[2]*Department of Earth Science & Engineering, Imperial College London, Royal School of Mines, Prince Consort Road, London, SW7 2AZ, UK*

[3]*GNSS Research Center, Wuhan University, Wuhan, China*

[4]*School of Geodesy and Geomatics, Wuhan University, 430079, Wuhan, China*

* Author to whom any correspondence should be addressed.

**Email**: dhaosgg@whu.edu.cn

## Abstract

Constraining the physical properties of Earth's deep interior, particularly the viscosity of the solid inner core and the density structure of large low-velocity provinces (LLVPs), remains a major challenge in geophysics. Here we develop a unified dynamical framework that combines mantle-inner core gravitational coupling (MICG) with torsional oscillations in the fluid outer core, and show that their interaction could produce a distinct and testable geodetic signature. Guided by this prediction, we analyze satellite gravity observations together with independent corrections for surface mass variability. We identify a robust ~6-year signal in the Stokes coefficient $\Delta S_{22}$, while no corresponding stationary signal is detected in $\Delta C_{22}$. A signal with the same periodicity is independently detected in length-of-day variations (ΔLOD), and the two signals exhibit a near anti-phase relationship. Interpreting this coupled signature within the proposed framework allows us to constrain the inner core's viscosity to $\sim 4.6(\pm 1.8)\times 10^{16}$ Pa·s and the equatorial relief of the inner core boundary to a semi-axis difference of about 200±70 m. The inversion further indicates mean density anomalies of +5.5(±0.6)‰ at the base of LLVPs. These results indicate that satellite gravimetry provides a direct observational window into deep-Earth dynamics and physical properties of Earth's deep interior.

## 1. Introduction

Understanding the dynamical processes operating within Earth's deep interior remains a central problem in geophysics. The core and the lowermost mantle play a key role in controlling the long-term evolution of the geodynamo, mantle convection, and Earth's rotational dynamics [1,2,3]. However, several fundamental physical properties of these regions, including the effective viscosity of the solid inner core [4,5,6,7] and the large-scale density structure of the lowermost mantle [8,9], remain poorly constrained by observations.

Most current knowledge of Earth's interior is derived from seismology. Seismic tomography has revealed two major large low-velocity provinces (LLVPs) beneath Africa and the Pacific, characterized by reduced seismic velocities, large spatial extent, and a dominant degree-2, order-2 equatorial antipodal configuration [10,11]. However, the conversion of seismic velocity anomalies into density and rheological parameters remains highly non-unique, leaving key quantities such as inner core viscosity and LLVP-related density contrasts poorly constrained [8,9,12,13].

Satellite geodesy provides an independent observational perspective. Measurements of Earth rotation, including length-of-day variations (ΔLOD), and low-degree gravity field coefficients, are sensitive to large-scale mass redistribution and angular momentum exchange within the Earth system. On interannual to decadal timescales, both ΔLOD [14,15] and the low-degree gravity coefficients (notably $\Delta C_{22}$ and $\Delta S_{22}$) show significant variability [16], suggesting a possible connection to dynamical processes in the core-mantle system [1,2,17,18].

Several mechanisms have been proposed to explain these signals. One class of models invokes mantle-inner core gravitational coupling (MICG), in which large-scale density heterogeneities in the lowermost mantle exert gravitational torques on the inner core and drive differential rotation [1,18,19]. Another class attributes ΔLOD to magnetohydrodynamic processes in the fluid outer core, especially torsional waves [20,21]. While both mechanisms can account for aspects of the observations [15,21,22], they are typically considered in isolation and rely on simplifying assumptions. As a result, neither mechanism alone can simultaneously explain both the rotational (ΔLOD) and gravitational ($\Delta C_{22}/\Delta S_{22}$) signatures in a dynamically consistent manner [18]. This limitation motivates the need for a unified dynamical framework that can jointly explain both types of observations within a physically consistent system.

In this study, we construct a dynamical framework by coupling mantle-inner core gravitational interaction with torsional dynamics in the fluid outer core. The mantle, fluid outer core, and solid inner core are treated as a coupled system, in which gravitational [1,19], electromagnetic [23,24], and topographic [25] interactions jointly control angular momentum exchange. Within this framework, the inner core exhibits a forced oscillatory response that depends on both fluid-core dynamics and inner core rheology. Although the amplitude of this motion is intrinsically small, it may produce a characteristic and potentially observable signal in the degree-2 order-2 components of the Earth's gravity field. The formulation of this framework is presented in Section 2.

Guided by this formulation, Section 3 analyzes long-term satellite gravimetry data with independent corrections for surface mass variability based on multiple hydrological models. We test whether a common signal exists in both ΔLOD and $\Delta C_{22}/\Delta S_{22}$. If confirmed and shown to be independent of external sources, such a signal provides direct observational evidence for inner core dynamics. Interpreting it within the proposed framework provides quantitative constraints on inner core viscosity and the effective density structure of the lowermost mantle (Section 4).

## 2. Theory and Prediction

To establish a physically consistent link between deep-Earth dynamics and observable geodetic signals, a viable framework must connect mantle heterogeneity, core dynamics, and measurable variations in both Earth rotation and the gravity field.

Although the sign of the LLVP density anomaly remains debated, its large-scale geometry, characterized by a degree-2, order-2 spherical harmonic $Y_{2,2}(\Omega)$ structure, is relatively well constrained [10]. Based on the S40RTS model [26], the symmetry axis $x'$ of the LLVPs can be estimated to deviate from the zero-longitude meridian by ~10° (Figure 1a). If LLVPs have positive density anomalies, this deviation angle is the angle $\alpha_0$ (as shown in Figure 1a); otherwise, $\alpha_0$ will differ from this deviation angle by 90°.

Regardless of whether the LLVPs represent positive or negative density anomalies, long-term equilibrium implies that the inner core develops an equatorial $Y_{2,2}(\Omega)$-type topography. The symmetry axis of this deformation is expected to align with that of the effective positive density structure in the lowermost mantle [15]. If the inner core undergoes a periodic oscillation in the equatorial plane, it can generate a time-varying gravitational signal controlled by this $Y_{2,2}(\Omega)$ geometry, which may be detectable in the temporal variations of the degree-2, order-2 Stokes coefficients $\Delta C_{22}$ and $\Delta S_{22}$ [17]. In addition, conservation of angular momentum requires that such inner core motion be accompanied by a corresponding oscillation in the mantle, expressed primarily as variations in ΔLOD [1,14,19].

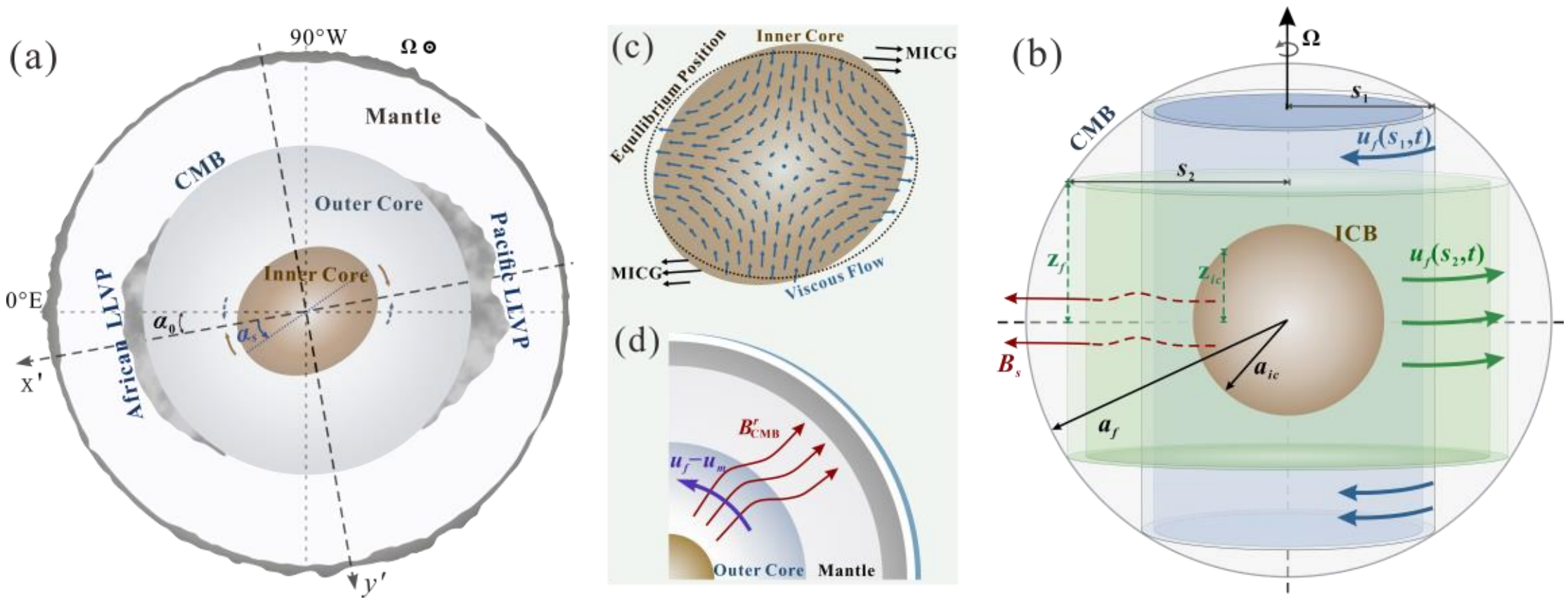

**Figure 1. (a) Inner core oscillation diagram:** The inner core oscillation is a periodic reversal relative to the mantle. The dashed black line is set to the $x'$-axis in the right-handed coordinate system, representing the symmetry axis of the LLVPs (~10°E-170°W) at the lower mantle. **(b) Schematic diagram of torsional oscillations in the outer core:** Cylindrical surfaces at different radii $s$ within the fluid core have different

angular velocities $u_f$. When differences in $u_f$ arise, they distort the magnetic field lines $B_s$ passing through the flow core, thereby generating a magnetic restoring force that induces torsional oscillations in the flow core. **(c) Schematic diagram of mantle-inner core gravitational coupling (MICG) and inner core viscous coupling:** When the inner core rotates away from its equilibrium state, it experiences MICG torque from the mantle, while simultaneously undergoing viscous flow (deformation) and rotational adjustment to realign with the mantle. **(d) Schematic diagram of core-mantle electromagnetic coupling:** When the angular velocities of the fluid core and the mantle (the inner core) differ, the magnetic field at the core-mantle boundary (CMB; the inner core boundary; ICB) is sheared, generating a magnetic restoring torque $f_m$ ($f_s$).

In the following, we shall theoretical analyze the MICG and torsional oscillation mechanisms, and explain that when considered them in isolation, neither MICG nor torsional oscillations can fully account for both the rotational and gravitational observations, nor can they provide robust constraints on key physical parameters. However, when these mechanisms are considered together within a coupled framework that better reflects the dynamics of the real Earth, such constraints become feasible.

### 2.1 Inner core and mantle rotational angles caused by the inner core oscillation

When the inner core exhibits a degree-2, order-2 spherical harmonic topography, i.e., an equatorially elliptic bulge, a longitudinal axial rotation by an angle $\alpha_{ic}$, as shown in Figure 1a, the change in the surface's gravitational field caused by the elastically misaligned inner core is [17,19,27]

$$\Delta U(a_e,\theta,\lambda) = i\frac{8\pi G\alpha_{ic}}{5a_e^3}\left[1+\tilde{k}_{\epsilon'}(a_e)\right](Y_{2,2}-Y_{2,-2})\cdot\int_0^{a_{ic}}(\rho-\rho_f')\frac{\mathrm{d}(a^5|\epsilon_{2,2}|)}{\mathrm{d}a}\mathrm{d}a \qquad (1)$$

where $G = 6.672\times10^{-11}$ N·m$^2$·kg$^{-2}$ denotes gravitational constant; $a_{ic} = 1.2215\times10^6$ m and $a_e$ = 6.371×10⁶ m denote the mean radius of the ICB and Earth, respectively; $\rho_f'$ represents density on the fluid core side of the ICB; the factor $[1+\tilde{k}_{\epsilon'}(a_e)]$ = 1.9736 accounts for the Earth's elastic deformations associated with the density change at the ICB [17]. $\epsilon_{2,2} = \epsilon_{2,2}(a)$ denotes the equatorial shape structure varying with mean radius $a$, and $Y_{2,2}(\theta, \varphi)$ is degree-2 order-2 fully normalized spherical harmonic function.

The corresponding degree-2 order-2 gravitational field Stokes coefficient change, and considering the difference between geographic coordinate and set in this study, it is written as

$$\begin{aligned}\Delta C_{22}+i\Delta S_{22} &= i\left[1+\tilde{k}_{\epsilon'}(a_e)\right]\frac{8\pi\sqrt{3}\xi_{ic}}{15\sqrt{5}Ma_e^2}\left[\int_0^{a_{ic}}(\rho-\rho_f{}')\frac{\mathrm{d}(a^5\overline{\epsilon})}{\mathrm{d}a}\mathrm{d}a\right]e^{-i2\alpha_0}\left(e^{i\alpha_{ic}}-1\right) \\ &\approx \upsilon\xi_{ic}\left[\sin(2\alpha_0)+i\cos(2\alpha_0)\right]\alpha_{ic}, \qquad \text{while } \alpha_{ic}\ll 1^\circ\end{aligned} \tag{2}$$

where $M$ = 5.974×10²⁴ kg is the total mass of the Earth; $\upsilon$ = 8.362×10⁻¹⁰ is derived from the PREM model [28]. The parameter $\xi_{ic} = (B_{ic}-A_{ic})/[C_{ic}-(A_{ic}+B_{ic})/2]$ denotes the triaxiality of the inner core (see Supplementary Materials for details), where $A_{ic}$ and $B_{ic}$ are the equatorial principal moments of inertia of the inner core, and $C_{ic}$ is the polar principal moment of inertia. The denominator depends solely on the flattening or geometrical ellipticity $\overline{\epsilon}$ (or also written $f$ in geodesy), and can therefore be directly evaluated. In contrast, the numerator $B_{ic}-A_{ic}$ remains unknown and corresponds to the degree-2, order-2 shape $\epsilon_{2,2}$. According to Eq. (2), even if $\alpha_0$ is known and the inner core oscillation together with its amplitude has been identified from the observed $\Delta C_{22}$ and $\Delta S_{22}$ coefficients, the inner core rotation angle $\alpha_{ic}$ still cannot be determined because $\xi_{ic}$ remains unknown. However, the product $\xi_{ic}\cdot\alpha_{ic}$ can be constrained. As will be shown below, this combined parameter is particularly informative when integrated with other geophysical observations.

The inner core oscillation through conservation of angular momentum, necessarily induce a compensating rotation of the mantle. The associated rotation angle can be obtained as

$$\alpha_m = \int \Delta\omega_m \mathrm{d}t = -\frac{\Omega_0}{\mathrm{LOD}} \int \Delta\mathrm{LOD}\mathrm{d}t \tag{3}$$

from the observed ΔLOD; where LOD = 86400s, and $\Omega_0$ denotes the mean sidereal rotation rate. Conversely, if an oscillatory signal is identified in ΔLOD, it is not possible to uniquely attribute it to inner core oscillation. Such variability may also arise from processes in the fluid core waves, inner core dynamics, or surface fluid mass redistribution (e.g., atmospheric, non-tidal oceanic, and hydrologic effects).

From the above analysis, it is evident that considering either ΔLOD or $\Delta C_{22}/\Delta S_{22}$ alone is insufficient to robustly identify inner core oscillation. Therefore, if a common oscillatory signal is detected in both datasets with consistent period and phase, the likelihood that it originates from inner core oscillation is significantly strengthened. Previous studies on periodic variations in ΔLOD have typically treated MICG and torsional oscillations in the fluid core as independent mechanisms. However, the Earth operates as a dynamically coupled system, in which these processes are inherently linked [29]. In the following, we attempt to jointly interpret these signals within a unified framework.

### 2.2 Earth's core motion under multi-mechanism coupling

Considering the coupling effects occurring at the ICB and CMB, the oscillation of inner core and torsional waves in the fluid core together form a coupled core torsional system [29,30]. The frequency domain form of the angular momentum equations governing the inner core, mantle, and outer core, respectively is written as [30]

$$\omega^2 C_{ic}\tilde{u}_{ic} + \hat{\Gamma}_z^{\omega}(\tilde{u}_m - \tilde{u}_{ic}) - i\omega\int_0^{a_{ic}} f_{ic}\mathrm{d}s = 0 \tag{4a}$$

$$\omega^2 C_m\tilde{u}_m - \hat{\Gamma}_z^{\omega}(\tilde{u}_m - \tilde{u}_{ic}) - i\omega\int_0^{a_f} f_m\mathrm{d}s = 0 \tag{4b}$$

$$\omega^2 4\pi\bar{\rho}_f s^3(z_f - z_{ic})\tilde{u}_f(s) + \frac{\mathrm{d}}{\mathrm{d}s}\left[4\pi s^3(z_f - z_{ic})\frac{\{B_s^2\}}{\mu_0}\frac{\mathrm{d}\tilde{u}_f(s)}{\mathrm{d}s}\right] + i\omega(f_{ic} + f_m) = 0 \tag{4c}$$

As there are too many parameters in Eq. (4), we listed them in Table 1. In Table 1, the relaxation time $t_\tau$ used for $\hat{\Gamma}_z^{\omega}$ can be estimated when assuming the viscosity $\eta$ of the inner core as a whole is low [31]

$$t_\tau = \frac{C_1\eta}{\Delta\rho_{\mathrm{ICB}}\, g_{\mathrm{ICB}}\, a_{ic}} \tag{5}$$

For the PREM model [28], the inner core boundary density jump $\Delta\rho_{\mathrm{ICB}} \approx 600$ kg·m$^{-3}$, and $g_{\mathrm{ICB}} \approx 4.4$ m·s$^{-2}$. $C_1$ =1.9 is a constant used by Ref. [31].

**Table 1. Parameters used for multi-mechanism coupling**

| Parameters | Definitions |
|---|---|
| $\omega = 2\pi/T - i/Q_t$ | The complex frequency; $T$ is period, $Q_t$ is decay time. |
| $\tilde{u}_{ic}$ and $\tilde{u}_m$ | Angular velocities of the inner core and the mantle |
| $\tilde{u}_f(s)$ | Angular velocity of fluid core at the cylindrical radius $s$ |
| $C_{ic}$ = 5.8673×10$^{34}$ kg·m$^2$ | Axial inertia moment of the inner core |
| $C_m$=7.1236×10$^{37}$ kg·m$^2$ | Axial inertia moment of the mantle |
| $a_{ic}$ = 1.2215×10$^6$ m | The mean radius of the ICB |
| $a_f$ = 3.48×10$^6$ m | The mean radius of the CMB |
| $\hat{\Gamma}_z^{\omega}(\omega) = \hat{\Gamma}_z\left(1 + i/\omega t_\tau\right)^{-1}$ | The frequency-dependent effective axial MICG constant, demonstrating that the shape of the inner core can makes viscous adjustments over the period of an oscillation (Fig. 1c). |
| $\hat{\Gamma}_z$ | The MICG constant in ideal state (inviscid inner core) |
| $-f_{ic}(s)$ and $-f_m(s)$ | The torques on the ends of the cylinders exerted by the inner core and mantle |
| | the half-height of the cylinders (see Fig. 1b); |
| $z_f - z_{ic}$ | $\begin{cases} z_f(s) = (a_f^2 - s^2)^{1/2},\ z_{ic}(s) = (a_{ic}^2 - s^2)^{1/2}, & s \le a_{ic} \\ z_f(s) = (a_f^2 - s^2)^{1/2},\ z_{ic}(s) = 0, & a_{ic} < s \le a_f \end{cases}$ |
| $\mu_0 = 4\pi\times10^{-7}$ Wb·(A·m)$^{-1}$ | The vacuum permeability |
| $\bar{\rho}_f = 1.1\times10^3$ kg·m$^{-3}$ | The mean density of the outer core |
| $\{B_s^2\}$ | The surface-averaged over the cylinder of radial magnetic field in the core |

As for the torques $f_{ic}(s)$ and $f_m(s)$, they depend on the coupling mechanism at the ICB and CMB. Specifically, considering the influence of relatively strong electromagnetic coupling (Fig. 1d), they are written as [24]

$$\begin{cases} f_{ic} = 2\pi s^3\left[(\hat{B}^r_{\mathrm{ICB}})^2 + (\bar{B}^r_{\mathrm{ICB}})^2\right] G_{ic}\left(\dfrac{a_{ic}}{z_{ic}}\right)[\tilde{u}_f(s) - \tilde{u}_{ic}] \\ f_m = 2\pi s^3\left[(\hat{B}^r_{\mathrm{CMB}})^2 + (\bar{B}^r_{\mathrm{CMB}})^2\right] G_m\left(\dfrac{a_f}{z_f}\right)[\tilde{u}_f(s) - \tilde{u}_{ic}] \end{cases} \tag{6}$$

where $\hat{B}^r_{\mathrm{ICB}}$ and $\hat{B}^r_{\mathrm{CMB}}$ denote the axial dipole component of the radial magnetic field across the ICB and CMB; $\bar{B}^r_{\mathrm{ICB}}$ and $\bar{B}^r_{\mathrm{CMB}}$ denote the RMS of the monopole component. The $G_{ic}$ and $G_m$ are the conductance at the ICB and CMB determined by [23,24]

$$\begin{cases} G_m = \displaystyle\int_{a_f}^{a_f+\Delta a_m} \sigma_m(r)\mathrm{d}r \\ G_{ic} = \dfrac{1}{4}(1 + i\dfrac{\omega}{|\omega|})\sigma_f\delta_f \end{cases} \tag{7}$$

where $\sigma_f = 5\times10^5$ S·m$^{-1}$ and $\sigma_m(r)$ are the conductivity of the core and conductivity profile of a thin layer of lower mantle basement, respectively; $\Delta a_m$ is the thickness of the layer. $\delta_f = 2/(2\omega\mu_0\sigma_f)^{1/2}$ is the magnetic skin depth, which depends on frequency and conductivity. Based on nutation observations, the electromagnetic coupling at the CMB can be explained by introducing a thin, electrically conducting layer in the lowermost mantle. The model requires that the thickness of this layer exceeds the magnetic skin depth associated with nutation, $\delta_{\text{nutation}}$ = ~210 m [32]. This condition imposes a lower bound on the integrated conductance, yielding $G_m \geq 10^8$ S. Previous studies have employed a conductance of $3\times10^8$ S by an infiltration layer 600 m thick [23], or a slightly lower conductivity but thick D″ layer [33]. Here, it is taken as $G_m = 3\pm2\times10^8$ S to include nutation, ΔLOD, and geomagnetic observations [23,32,34].

We further quantified the effect of topographic coupling at the CMB. Figure S1 shows 6 different CMB topographic models. Typically, the topographic coupling constant reaches the order of $10^{17}$ N·m (for example, the Morelli model [35]), compared with $10^{23}$ N·m near the equator for electromagnetic coupling (Figure S2), indicating that topographic coupling is weaker by 6 orders of magnitude. Hence, the topographic coupling contribution is found to be substantially smaller than that of electromagnetic coupling and can therefore be neglected (Figure S2).

Solving Eq. (4) needs some boundary conditions. First is that $\mathrm{d}\tilde{u}_f / \mathrm{d}s = 0$ at $s$ = 0, and due to the singular behavior of the electromagnetic coupling constant, we need constrain $\tilde{u}_f = \tilde{u}_m$ at $s = a_f$. By applying the boundary constraint and discretizing the governing equations (Eq. (4)), a matrix representation of the system can be constructed for numerical solution,

$$\mathbf{M}(\omega)\begin{bmatrix} \tilde{u}_{ic} \\ \tilde{u}_f(s_1) \\ \vdots \\ \tilde{u}_f(s_N) \end{bmatrix} = 0 \tag{8}$$

Eigen-frequencies $\omega$ and their corresponding eigen-functions $\mathbf{u} = [\tilde{u}_{ic}\ \tilde{u}_f(s_1)\ ...\ \tilde{u}_f(s_N)]^{\mathrm{T}}$ are obtained by numerical solution. For a specific torsional eigen-mode, the ratio of the angular rotations between the inner core and the mantle equals the ratio of their angular velocities

$$\zeta = \frac{\alpha_{ic}}{\alpha_m} = \left(1 + \frac{i}{\omega t_\tau}\right)\frac{\tilde{u}_{ic}}{\tilde{u}_N} \tag{9}$$

The real part of $\zeta$ indicates that the motion of the inner core and the mantle is either in phase or in antiphase, whereas the imaginary part represents the quadrature component (90° phase difference).

If only MICG is considered, i.e., retaining only the first two equations in Eq. (4), the system reduces to

$$\begin{cases} \omega^2 C_{ic}\tilde{u}_{ic} + \hat{\Gamma}_z(\tilde{u}_m - \tilde{u}_{ic}) = 0 \\ \omega^2 C_m \tilde{u}_m - \hat{\Gamma}_z(\tilde{u}_m - \tilde{u}_{ic}) = 0 \end{cases} \tag{10}$$

The corresponding eigen-period of the MICG mode, together with the ratio between the rotation angles of the inner core and the mantle, is then given by

$$\begin{cases} T = \dfrac{2\pi}{\omega} = 2\pi / \sqrt{\hat{\Gamma}_z (C_m + C_{ic}) / (C_m C_{ic})} \approx 2\pi / \sqrt{\hat{\Gamma}_z / C_{ic}} \\ \dfrac{\alpha_{ic}}{\alpha_m} = -\dfrac{C_m}{C_{ic}} \end{cases} \tag{11}$$

This behavior differs fundamentally from that described by Eq. (9). In this case, the eigen-period depends solely on the unknown MICG constant $\hat{\Gamma}_z$, while the ratio $\alpha_{ic}/\alpha_m$ remains fixed and is independent of the viscosity of the inner core.

When only torsional oscillations are considered, the third term in Eq. (4) reduces to

$$\omega^2 4\pi\bar{\rho}_f s^3 z_f \tilde{u}_f(s) + \frac{\mathrm{d}}{\mathrm{d}s}\left[4\pi s^3 z_f \frac{\{B_s^2\}}{\mu_0}\frac{\mathrm{d}\tilde{u}_f(s)}{\mathrm{d}s}\right] = 0 \tag{12}$$

In previous studies, the magnetic tension $4\pi s^3 z_f \{B_s^2\}/\mu_0$ is assumed to be independent of the cylindrical radius $s$, reducing the system to a one-dimensional eigenvalue problem. The corresponding oscillation period $T \approx a_f / \sqrt{\{B_s^2\}/(\bar{\rho}_f \mu_0)}$ is then controlled by the internal magnetic field strength and distribution, and is theoretically independent of both $\alpha_{ic}$ and $\alpha_m$.

### 2.3 The effects of the density of the mantle on the coupling mechanism

If density anomaly of the mantle, particularly the anomaly near the CMB, and the equatorial shape of the inner core are known, the MICG constant $\hat{\Gamma}_z$ can be calculated by the degree-2, order-2 density multipoles, expressed as [19,36,37]

$$\hat{\Gamma}_z = \frac{32\pi}{5} G\gamma f_s \left|q_{22}^m\right| \left|Q_{22}^{ic}\right| \tag{13}$$

where the factor $\gamma = \Delta\rho_{\mathrm{ICB}}/\rho_{ic} = 0.0468$ accounts for the hydrostatic pressure effect of the passive outer core, and the factor $f_s = 0.85$ account for elastic deformation of the inner core oscillation [36,38]. The degree-2 order-2 multipole of the exterior type belonging to the inner core $Q_{22}^{ic}$ is written as [19,37]

$$\begin{aligned} \left|Q_{22}^{ic}\right| &= \iiint_{ic} \rho(\mathbf{r}) r^4 Y_{2,2}^*(\Omega) \mathrm{d}r \mathrm{d}\Omega = \rho_{ic} a_{ic}{}^5 \left|\epsilon_{2,2}^{\mathrm{ICB}}\right| \\ &= \sqrt{\frac{15}{32\pi}} (B_{ic} - A_{ic}) = \frac{1}{8}\sqrt{\frac{15}{2\pi}} (A_{ic} + B_{ic}) e_{ic} \xi_{ic} \end{aligned} \tag{14}$$

where $e_{ic} = 2.422\times10^{-3}$ denotes the dynamical ellipticity of the inner core [28]; $A_{ic}$ and $B_{ic}$ are the equatorial principal moments of inertia of the inner core, and the mean equatorial moment is given by $(A_{ic}+B_{ic})/2 = C_{ic}/(1+e_{ic})$.

In the core environment, if the shape of Earth's core is induced by mantle density anomalies, the triaxiality of the inner core can be directly related to the degree-2, order-2 internal density multipole of mantle, which is given in the Supplementary Materials and serves as a measure of mantle density heterogeneity. Under this framework, the same multipole can also be used to infer the corresponding degree-2, order-2 topography of the CMB.

If one considers only a localized density anomaly of ~+1% at the base of the LLVPs, it would exert a downward load on the CMB, implying a local depression of the boundary beneath the LLVPs. However, when the entire mantle is taken into account, together with the fact that the density on the fluid core side of the CMB is approximately twice that on the mantle side, as well as self-gravitational effect, the CMB is expected to exhibit an overall topographic uplift under long-term dynamical equilibrium (see Supplementary Materials for detail). In the case where the LLVPs are characterized by negative density anomalies, the orientation of the CMB topography is reversed. The following discussion is therefore conducted within the framework of the Earth's long term equilibrium state, accounting for the entire mantle. Thus, the degree-2, order-2 shape of isopycnic surfaces satisfies the Clairaut equation [39] and is primarily controlled by lateral density variations in the mantle [39]. This allows the mantle internal degree-2, order-2 density multipole to be directly linked to the geometric parameters of the inner core,

$$q_{22}^{m} = \frac{5\epsilon_{2,2}^{\mathrm{CMB}}}{a_f^3}\int_0^{a_f}\rho(a)a^2\mathrm{d}a - \frac{1}{a_f^5}\int_0^{a_f}\rho(a)\frac{\mathrm{d}(a^5\epsilon_{2,2})}{\mathrm{d}a}\mathrm{d}a \equiv o_1\xi_{ic} \tag{15}$$

Based on the PREM model in the Supplementary Materials, $o_1$ = 12.606 kg·m$^{-3}$ can be obtained.

This multi-mechanism coupling frame establishes a direct link between deep-Earth structure, core dynamics, and observable geodetic signals, forming the basis for the joint constraints developed in the following sections.

## 3. Results from ΔLOD and ($\Delta C_{22}/\Delta S_{22}$) observations

The primary objective of this section is to identify signals with common periodicities in ΔLOD and ($\Delta C_{22}/\Delta S_{22}$), and to analyze their phase and amplitude characteristics, in order to more robustly assess whether these signals may originate from core dynamics. Since satellite-derived observations of ($\Delta C_{22}$) and ($\Delta S_{22}$) are available from 2002 to the present, our analysis is restricted to periods shorter than 10 years.

### 3.1 Estimating the misalignment angle from ΔLOD

The fluctuations in the ΔLOD (1962/1-2020/2) were meticulously corrected using external sources (atmospheric and oceanic effects; Figure S5a), and the hydrological effect is not further considered due to the fact that it is far below even the background noise of ΔLOD in the target frequency band (Figure S5b and Supplementary Text). Subsequent removal of tidal effects based on a tidal model [40], and after further employing a 6-month running mean to reduce significant high-frequency noises, a smoothed ΔLOD sequence $g_1(t)$ is obtained (Figure 2a). To avoid possible low-frequency leakage associated with band-pass filtering, we fitted and removed from ($g_1(t)$) the previously reported oscillatory components with periods longer than 10 years [41], and defined the residual as $R(t)$. The amplitude of the long-period variability retained in $R(t)$ is about 0.2 ms. The most significant period in $R(t)$ and $\alpha_m(t)$ is ~6-year (see Figure S5); this finding is consistent with those of previous studies [22,41].

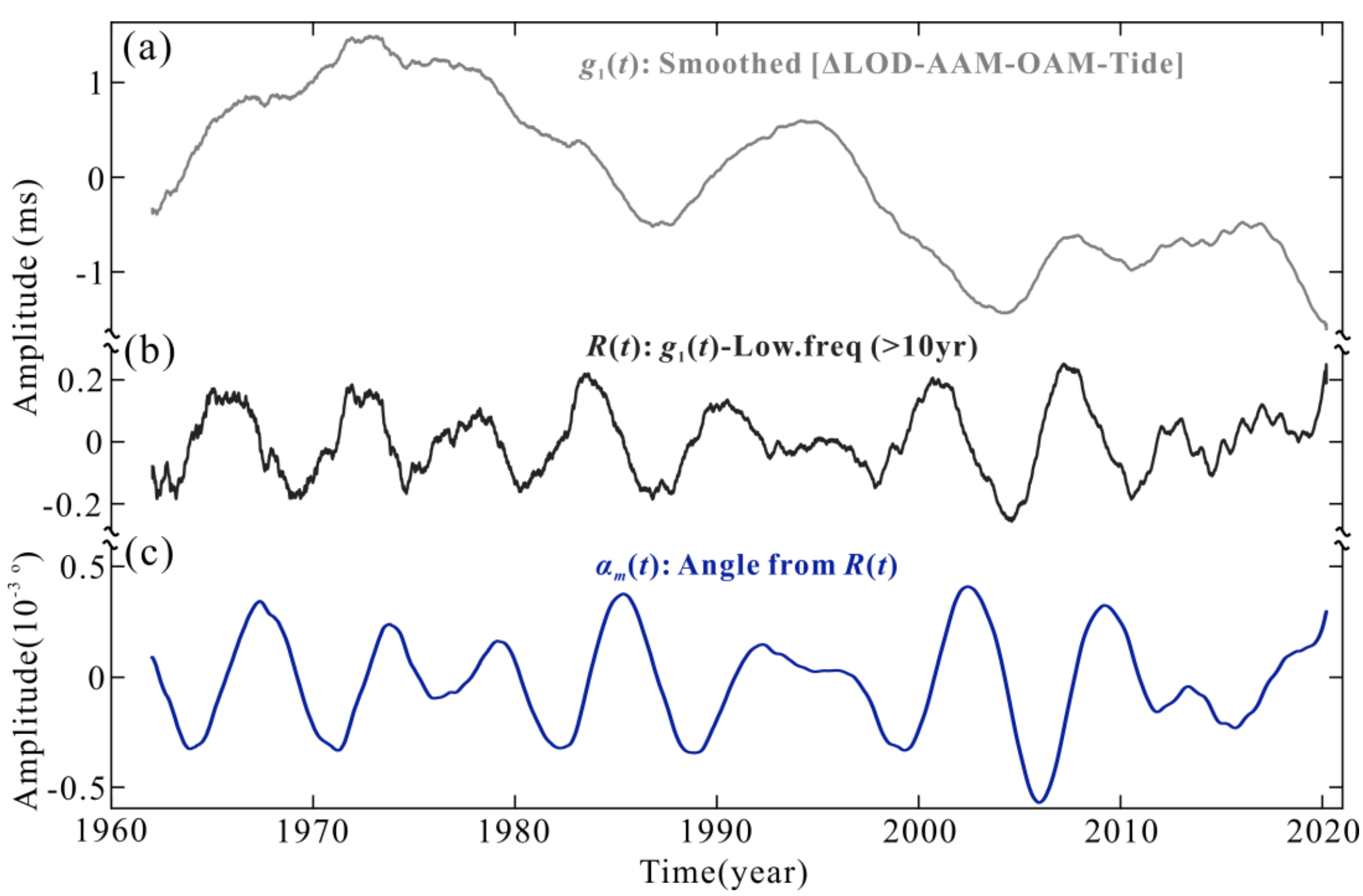

**Figure 2: Angle $\alpha_m(t)$ of mantle motion associated with inner core motion inferred from the ΔLOD time series.** (a) The smoothed ΔLOD record $g_1(t)$ with atmospheric and oceanic effects and tide removed, and a 6-month running mean window used. (b) Residual sequence $R(t)$ in which low-frequency signals with a period of more than 10-year are removed from $g_1(t)$ sequences. (c) Angular progression of mantle rotation $\alpha_m(t)$ inferred from $R(t)$.

### 3.2 The angle of inner core motion $\alpha_{ic}(t)$ from $\Delta C_{22}$ and $\Delta S_{22}$

The $\Delta C_{22}$ and $\Delta S_{22}$ sequences (2002/1-2024/1) determined by SLR (satellite laser ranging) from CSR (Center for Space Research, University of Texas at Austin; recognized products) are firstly corrected using external sources (atmospheric, oceanic, and hydrological effects; AOH effects). As shown in the Fourier power spectra of the $\Delta C_{22}$ and $\Delta S_{22}$ sequences (Figure 3a), there is only a significant peak corresponding to a ~6-year period in the spectrum of the $\Delta S_{22}$ sequence for the 2-20 period band, while for the $\Delta C_{22}$ sequence, the significant spectral peak located in the period range of 12-20 years, and a much weaker spectral peak with ~5.3-year period. As the length of the used $\Delta C_{22}$ is only 22 years, the 12-20 year spectral peak is unlikely to represent a stationary oscillation; rather, it may reflect the superposition of multiple periodic components that cannot be reliably separated because of the limited data length. Given that Fourier spectra are susceptible to non-stationary signals and thus to false results, we re-analyzed these two sequences in Figure 3b by using the stabilized AR-z methods; this method helps to identify stationary harmonics [42] and has higher frequency resolution than Fourier spectra [43]. Figure 3b clearly shows a robust ~6-year signal in $\Delta S_{22}$ but no corresponding signal in $\Delta C_{22}$. Besides, no significant spectral peak in the period of 12-20 years, which means that the spectral peak in the period of 12-20-year identified in Figure 3a is not a stationary signal.

We further analyzed the other $\Delta C_{22}$ and $\Delta S_{22}$ sequences from different sources (Figure S6), the related results also show that only a ~6-year period signal can be stably detected in only $\Delta S_{22}$. To verify those results, we also show the Morlet wavelet spectra of the used $\Delta S_{22}/\Delta C_{22}$ (Figure S7 and Supplementary Text); results also show that the ~6-year signal is only present in $\Delta S_{22}$, and limited by period-resolution of the Morlet wavelet spectrum, the 12-20 years oscillation still presents in $\Delta C_{22}$ but without a near-fixed period. Those findings indicate that the ~6-year signal identified in $\Delta S_{22}$ is robust. According to Figure 3, the amplitude of the ~6-year signal in $\Delta S_{22}$ (hereafter also referred to as the six-year oscillation, SYO) is estimated to be $\sim 1.12\times10^{-11}$. An independent $\Delta S_{22}$ dataset analyzed in the Supplementary Materials yields a consistent estimate of $\sim 1.04\times10^{-11}$. Combining these results, we adopt a weighted amplitude of $1.08(\pm0.05)\times10^{-11}$.

A similar ~6-year signal may also be present in $\Delta C_{22}$, but its amplitude appears to be smaller not only than that of the corresponding signal in $\Delta S_{22}$, but also than the background noise level, making it difficult to detect. Given that the background noise level in $\Delta C_{22}$ is about $4.5\times10^{-12}$, we infer that even if the SYO is present in $\Delta C_{22}$, its amplitude is likely smaller than $4.5\times10^{-12}$. This implies that the amplitude ratio of the SYO in $\Delta S_{22}$ to that in $\Delta C_{22}$ should exceed $1.08/0.45 \approx 2.4$. If the SYO in $\Delta S_{22}$ originates from inner core motion, then, according to Eq. (2), it is linearly proportional to $\alpha_{ic}(t)$ (with the proportionality constant involving the unknown $\xi_{ic}$). Therefore, $\alpha_{ic}(t)$ must also contain the SYO.

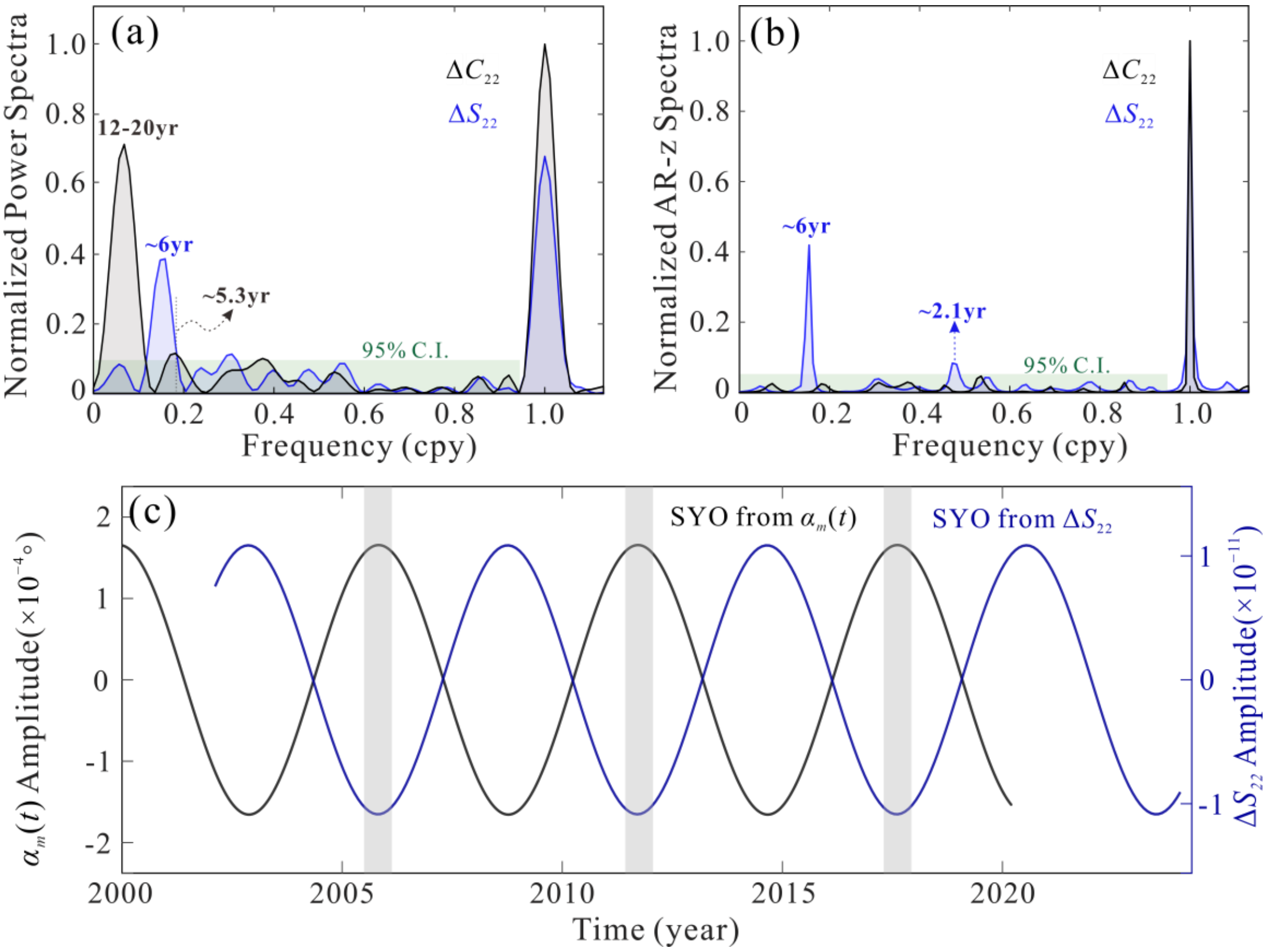

**Figure 3: Verification of the six-year oscillation (SYO) for the inner core oscillation from the length-of-day variation (ΔLOD) and degree-2 order-2 Stokes coefficients.** (a) The normalized Fourier power spectra of the corrected degree-2 order-2 Stokes coefficient variations $\Delta C_{22}$ and $\Delta S_{22}$ sequences (2002/1-2024/1) determined by SLR;

green region represents the 95% confidence interval (C.I). (b) Similar to (a) but for the stable AR-z spectra. (c) Fitted SYOs from corrected $\Delta S_{22}$ sequences and angle $\alpha_m(t)$ using the classical least-squares method, respectively.

### 3.3 Identification of the inner core motion signal

Based on the results of Sections 3.1 and 3.2, we identify a common SYO signal in both ΔLOD (i.e., $\alpha_m(t)$) and $\Delta S_{22}$ (linked to $\alpha_{ic}(t)$). Since the $\Delta S_{22}$ result is obtained after removing external contributions, and we have independently verified that these external effects do not contain a stable ~6-year signal (Figure S8), it is reasonable to infer that the SYO in $\Delta S_{22}$ originates from inner core motion. By the same argument, the SYO identified in ΔLOD is also likely associated with core dynamics. To further establish that the two signals are of common origin and indeed reflect core motion, the theoretical framework introduced in Section 2.1 requires that the oscillation angles $\alpha_m(t)$ and $\alpha_{ic}(t)$ not only share the same period but also exhibit either the same phase or an opposite phase relationship. Figure 3c compares the SYO extracted from $\alpha_m(t)$ with that extracted from $\Delta S_{22}$ (i.e., $\upsilon\cos(2\alpha_0)\xi_{ic}\alpha_{ic}(t)$). The two signals are found to be in opposite phase. This phase relationship therefore provides stronger evidence that the SYOs identified in ΔLOD and $\Delta S_{22}$ represent a common signal associated with inner core motion.

## 4. Inversions for physical parameters in the Earth's deep interior

Since we have established with reasonable robustness that the SYO identified in both ΔLOD and $\Delta S_{22}$ originates from inner core motion, the theoretical framework developed in Section 2 makes it possible to further constrain the viscosity of the inner core and the strength of MICG, and thereby to place additional constraints on the density anomaly of LLVPs.

According to Section 2.1, the amplitude ratio between the signals in $\Delta S_{22}$ and $\Delta C_{22}$ is $\cos(2\alpha_0)/\sin(2\alpha_0)$. From Figure 3, this ratio is constrained to be larger than 2.4, which requires $\alpha_0 \leq 11.3°$. This value is close to the estimated upper bound of ~10° the deviation angle of the shape axis $x'$ of the LLVPs that we have estimated. Therefore, we may suggest that the LLVPs should have a positive relative density, and suggest $\alpha_0 \approx 10°$ as a reference.

We also confirmed that the two fitted SYOs from $\alpha_m(t)$ and $\alpha_{ic}(t)$ have nearly opposite phases. This finding implies that the inner core oscillation is in antiphase with the mantle motion, with almost no quadrature component (90° phase difference). Within the coupled framework of Section 2.2, this requires $\mathrm{Re}(\zeta) >> \mathrm{Im}(\zeta)$. Moreover, combining the observed relation $\Delta S_{22}/\alpha_m = -6.558\times10^{-8}$ (Figure 3) and $\Delta S_{22} = \upsilon\cos(2\alpha_0)\xi_{ic}\alpha_{ic}$ ($\alpha_{ic}$ in degrees), we can obtain $\xi_{ic}\alpha_{ic}/\alpha_m = \xi_{ic}\zeta = -83.45$ ($\upsilon = 8.362\times10^{-10}$ based on PREM model [28]). This ratio and $\mathrm{Re}(\zeta) >> \mathrm{Im}(\zeta)$ are very useful for constraining physical parameters of the Earth's deep interior based on physical numerical simulations.

In the coupled Earth's core torsional system explained in Section 2.2, according to Eqs. (13), (14), and (15), the relationship between the triaxiality parameters of the inner core and MICG constants is $\hat{\Gamma}_z \approx 3.7016\times10^{22}\xi_{ic}{}^2$ as illustrated by the blue curve in Figure 4a. This inverse relationship corresponds to the measured constraint we get. Note that the ratio $\zeta$ is not only controlled by $\hat{\Gamma}_z$ but also controlled by the viscosity $\eta$ of the inner core. Referred to the *gufm* field model, the RMS of axial dipole component and nondipole component of radial magnetic field across the CMB are set to 0.226 mT and 0.42 mT [44], respectively. Similarly, they are set to 2 mT and 3 mT across the ICB, respectively [24]. We can use the complex coupling mechanism framework constructed in Section 2.2 for numerical modeling. The magnetic field $\{B_s^2\}$ is set to ~2 mT [21], and the eigen-period of the fundamental mode of the system roughly corresponds to the intradecadal scale. By jointly varying the inner core viscous relaxation time $t_\tau$ and the MICG constant $\hat{\Gamma}_z$, we require that the modeled eigen-period matches the observed ~6-year period, while also satisfying the observed inverse relation between the relevant model parameter and $\zeta$ (blue curve in Figure 4a), together with the condition $\mathrm{Re}(\zeta) \gg \mathrm{Im}(\zeta)$. One representative fit is shown in Figure S9, yielding $t_\tau$ = 0.682 years and an MICG constant of $1.8\times10^{20}$ N ·m. Extending this procedure across the admissible parameter space gives an MICG constant of ~0.7-3.1×$10^{20}$ N ·m, corresponding to $t_\tau$ ~ 0.5-1.2 years and an average inner core viscosity of ~2.8-6.4×$10^{16}$ Pa· s. These ranges include the amplitude uncertainties of the SYO inferred from both $\Delta S_{22}$ and $\alpha_m(t)$.

Our simulation results also indicate that the difference between the equatorial semi-axes of the inner core is about $\Delta h_{\mathrm{ICB}}$ = 200±70 m, and further constrain the range of the mantle's degree-2 order-2 internal density multipole $q_{22}^m$ to 0.85±0.3 kg·m$^{-3}$, which is mainly contributed to both of the CMB topography and the density anomaly of LLVPs with a degree-2 order-2 spherical harmonic $Y_{2,2}(\Omega)$ spatial structure in the lower mantle. As discussed above, according to several published models (Figure S1), the gravitational forcing of mantle density heterogeneities causes the CMB topography to an approximate positive $Y_{2,2}(\Omega)$ shape (see Supplementary Materials). Given the large deviations among different models, we do not intend to adopt any of them. Following [39], $\epsilon_{2,2}^{\mathrm{CMB}}$ exceeds $\epsilon_{2,2}^{\mathrm{ICB}}$ by about 5% (see Figure S4), corresponding to a CMB topographic amplitude of $\Delta h_{\mathrm{CMB}}$ = ~600±210 m in the equatorial plane, which is broadly consistent with some previously proposed models (see Figure S1). Hence, the topographical effect of the CMB $q_{22}^{\mathrm{CMB}}(= \Delta\rho_{\mathrm{CMB}}\,\epsilon_{2,2}^{\mathrm{CMB}})$ can be obtained as 0.48±0.17 $\mathrm{kg{\cdot}m^{-3}}$.

After correcting for CMB topography, the degree-2, order-2 internal density multipole $q_{22}^{m}$ is estimated to be ~0.37±0.13 $\mathrm{kg{\cdot}m^{-3}}$, which is primarily attributed to lateral density heterogeneity in the mantle. According to its definition, the multipole can be interpreted as a weighted radial integral of the mantle density anomaly with a $Y_{2,2}(\Omega)$ spatial structure, i.e., $\varepsilon_{22}^{*}(a)$, with weighting function $p(a)(= \rho(a)/a)$ (see Supplementary Materials). As shown in Figure S3, the weighting function $p(a)$ decreases with increasing mean radius $a$, indicating that the multipole is most sensitive to density anomalies in the lowermost mantle, particularly near the CMB. Therefore, the present inversion mainly constrains the density anomaly at the base of the LLVPs.

Constraints on density variations in other parts of the mantle are derived primarily from seismic shear-wave velocity ($v_s$) models using a logarithmic conversion factor, defined as $R_\rho = \partial\ln\rho/\partial\ln v_s$. For the upper mantle and the upper portion of the lower mantle (depth 0-1801 km), we adopt the scaling factors $R_\rho$ from Refs. [13,45]. In the mid- to lower part of the lower mantle (depth 1801-2891 km), different seismic tomography models consistently show that the LLVPs exhibit similar long-wavelength structures [13], particularly in the degree-2, order-2 spatial pattern. Sensitivity analyses further demonstrate that tidal observations of the $M_2$ semi-diurnal body tide, which are also dominated by degree-2, order-2 components, exhibit a strong response to LLVP-related density structures [13], particularly when adopting the S40RTS model [45]. For this reason, we primarily discuss results based on the S40RTS model for comparison. For the upper (1801-2141 km) and middle (2141-2591 km) portions of the LLVPs, the conversion factors are set to $R_\rho = 0.2$ and 0.15, respectively [45]. Given that a primordial chemical reservoir may exist at the base of the LLVPs [46], we assume a uniform $R_\rho$ from the CMB down to 2591 km depth. This depth interval is the focus of our inversion, and $R_\rho$ within this range is treated as the primary adjustable parameter. By adjusting $R_\rho$, applying crustal corrections using the CRUST1.0 model [47], and enforcing the constraint from the internal density multipole $q_{22}^m$, we obtain the density anomaly at the base of the LLVPs. Figure 4b shows the resulting density anomaly distributions at depths of 2891 km, 2731 km, and 2591 km (using the previously derived $q_{22}^m$ = ~0.37 kg·m$^{-3}$ as reference). At the base of the mantle, the maximum density anomaly reaches ~+1.5%, with an average of ~+6‰. At depths of 2731 km and 2591 km, although the spatial distributions differ, the peak and mean values remain similar, with maxima of ~+1% and averages of ~+5‰.

Figure S10 also present results based on the S362ANI model [48], which yields a more laterally uniform structure with an average density anomaly of ~+5‰, a maximum of ~+1.5% at the base, and values up to ~+1% at 2591 km depth. These results are broadly consistent in magnitude with those obtained using the S40RTS model. Overall, we favor the results derived from the S40RTS model.

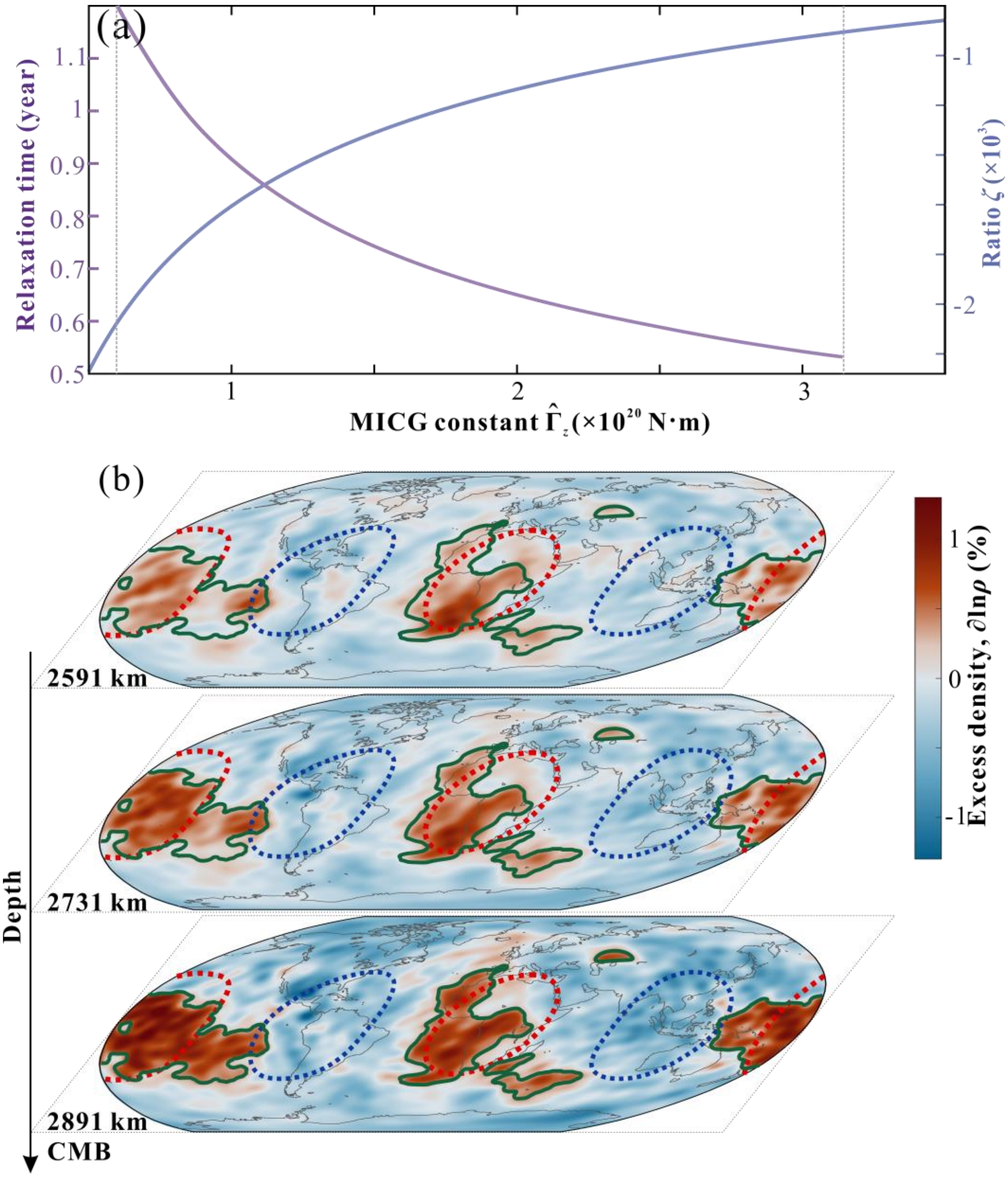

**Figure 4: The inversion of physical parameters in the Earth's deep interior.** (a) The curve of the angular amplitude ratio $\zeta$ between the inner core and the mantle as a function of the MICG constant $\hat{\Gamma}_z$, derived from observations, together with the curve of the inner core viscous relaxation time as a function of the MICG constant obtained by fitting the observational curve through numerical simulations; dashed lines represent the boundaries of fitted MICG constant $\hat{\Gamma}_z$. (b) The excess density $\partial \ln\rho$ (in %) converted by the S40RTS model at depths of 2891 km, 2731 km, and 2591 km, is based on the mantle's interior multipole value of 0.37 kg·m$^{-3}$; the green curves circle the contour of the LLVPs defined by the −6.5‰$V_s$; the dashed curves circle the sensitive region of degree-2 order-2 density multipoles and different colors mean that they have opposite signs.

## 5 Discussion and Conclusion

Motions of Earth's core are inherently difficult to observe directly. A variety of observations, including seismic waves, Earth's rotation, and core surface flow, have suggested core dynamics spanning timescales from hours to decades, although many of these signals remain debated [41,49,50]. As a dynamically coupled system, the Earth is expected to exhibit coherent responses across its internal layers when motions originate in the inner or outer core.

In this study, building upon previously proposed MICG and torsional oscillation mechanisms for decadal core dynamics, we construct a unified framework that explicitly incorporates the coupled dynamics of the inner core, outer core, and lowermost mantle. In addition to electromagnetic coupling, topographic coupling at the CMB is also evaluated. However, given that its contribution is ~6 orders of magnitude smaller than electromagnetic coupling, it is neglected in the final analysis. Within this framework, we derive the theoretical responses of low-degree gravity field Stokes coefficients and ΔLOD, and demonstrate how these observables can be used to constrain key deep-Earth parameters, including inner core viscosity and the base density structure of the LLVPs.

We first recover a robust ~6-year periodic signal in ΔLOD and exclude atmospheric, non-tidal oceanic, and hydrological contributions as dominant sources. This confirms a core origin for the signal, although it does not uniquely distinguish between inner and outer core processes. We then identify a corresponding ~6-year signal in the satellite-derived gravitational coefficient $\Delta S_{22}$, while no significant signal is detected in $\Delta C_{22}$. After excluding external excitations, we attribute the $\Delta S_{22}$ signal primarily to inner core motion. Individually, however, each observable provides only limited constraints: ΔLOD alone cannot uniquely determine the underlying mechanism, while $\Delta S_{22}$ depends on poorly constrained inner core triaxial parameters.

A key result of this study is that the joint analysis of ΔLOD and $\Delta S_{22}$ within the coupled framework significantly reduces these ambiguities. The phase coherence of the ~6-year signals in both observables indicates a common origin associated with inner core oscillation, coupled with outer core dynamics. In addition, the absence of a detectable signal in $\Delta C_{22}$ constrains the orientation of the inner core symmetry axis, with $\alpha_0$ inferred to be less than 11.3°, consistent with independent seismic inferences of LLVP geometry. These constraints lead to a robust proportional relation, $\xi_{ic}\cdot\zeta = -83.45$, which enables inversion for both inner core viscosity and LLVP density anomalies.

Our inversion yields an inner core viscosity of $\eta \approx 4.6(\pm 1.8)\times 10^{16}$ Pa· s, consistent with the lower bound of recent geodynamical and mineral physics estimates [4,5,7]. Importantly, this viscosity range does not arise from fitting a single observable, but emerges from the coupled dynamical constraints imposed by multiple independent observations. In particular, the admissible parameter space is jointly restricted by requiring that the model simultaneously reproduces the observed ~6-year eigen-period, the amplitude relationship between ΔLOD and $\Delta S_{22}$, and their temporal consistency within the coupled mantle-outer core-inner core system. These constraints are intrinsically coupled and cannot be satisfied independently in simplified models, such that only a limited range of viscosity values allows the system to match all observations simultaneously.

From a physical perspective, this viscosity implies that the inner core operates in a viscoelastic regime on intradecadal timescales. The corresponding viscous relaxation time (~0.5-1.2 years) is shorter than, but comparable to, the oscillation period (~6 years), indicating that the inner core can undergo substantial internal adjustment during each oscillation cycle while maintaining a coherent large-scale structure. This allows the inner core to both participate in angular momentum exchange with the mantle and fluid core and dissipate energy through internal deformation, suggesting a transitional rheological behavior between purely elastic and purely viscous responses.

We now turn to the implications for the density structure of the lowermost mantle. Our results indicate a depth-dependent LLVP density anomaly, increasing toward the CMB from approximately +(4.8±0.5)‰ at ~300 km above the CMB to +(6.4±0.7)‰ at the CMB. This estimate primarily reflects the degree-2, order-2 component of the density field, which is robust across different seismic tomography models despite variations in smaller-scale structures. This suggests that the inferred density anomaly is not strongly dependent on the specific velocity model adopted, but instead reflects a large-scale feature of the lowermost mantle.

As in comparable studies, uncertainties remain. These arise primarily from two sources. First, observational uncertainties are dominated by the estimation of $\alpha_0$. Through comparisons using different datasets and external fluid models, we find that uncertainties in the amplitude of the ~6-year signal in $\Delta S_{22}$ have a relatively minor impact (~2%). Although the weak signal in $\Delta C_{22}$ introduces larger uncertainty, independent constraints from LLVP geometry and prior studies support a positive density anomaly, leading us to estimate that the uncertainty in $\alpha_0$ is within ~2°. This propagates into uncertainties in $\eta$, $\Delta h_{\mathrm{ICB}}$, and LLVPs anomalous density of approximately 2%, 0.5%, 0.2%. Second, model-related uncertainties arise from assumptions in the coupling framework, including inner core magnetic field strength, mantle 3D structure, and other coupling parameters. Based on more than one hundred synthetic experiments, we estimate their contributions to uncertainties in $\eta$, $\Delta h_{\mathrm{ICB}}$, and LLVPs anomalous density to be about half of the uncertainties caused by the used $\alpha_0$. In addition, the uncertainties of the used $\Delta\rho_{\mathrm{ICB}}$ ($\approx$ 600±60 kg·m$^{-3}$) and $C_1$ ($\approx$ 1.9±0.1) in Eq. (5) also introduced some errors, but mainly for $\eta$; the corresponding error is about $7\times10^{15}$ Pa·s, which is within the reported range of $\pm1.8\times10^{16}$ Pa·s. These effects are incorporated into the reported error bounds.

Overall, this study demonstrates that integrating multiple observational datasets provides a more robust assessment of core dynamics than approaches relying on a single observable or mechanism. Our results highlight the potential of satellite geodesy to place quantitative constraints on the rheology, geometry, and density heterogeneity of the deep Earth. The combined use of gravity observations, Earth rotation data, and coupled dynamical modeling offers a promising pathway toward probing the physical state and long-period dynamics of Earth's deepest interior.

## Data availability

The ΔLOD and external sources datasets used in this work are available via IERS (https://www.iers.org/IERS/EN/DataProducts/EarthOrientationData/eop.html). The degree-2 order-2 gravitational field Stokes coefficients measured by the SLR and GRACE can be respectively downloaded from https://filedrop.csr.utexas.edu/pub/slr/degree_2/C22_S22_RL06.txt and https://icgem.gfz-potsdam.de/sl/temporal. Hydrological model is from GES DISC (nasa.gov). The mantle's speed models S40RTS is from SubMachine: Web-based tools for exploring seismic tomography and other models of Earth's deep interior (ox.ac.uk).

## Code availability

The test code of the AR-z spectrum has been uploaded to https://agupubs.onlinelibrary.wiley.com/doi/10.1029/2018JB015890. It is also available from the corresponding author H.D upon request.

## Acknowledgments

This study is supported by the National Natural Science Foundation of China (Grants 42388101, 42388102 and 42404003), Natural Science Foundation of Wuhan (Grants

2024040701010027), and Natural Science Foundation of Hubei Province (Grants 2024AFB143).

**Authors contributions**

H.D, W.J and J.L supervised the project. Y.A conceived the idea, designed the experiments, conducted the data analysis, comparison, modelling, and wrote the first draft. H.D conducted the data analysis, prepared the figures, assisted in the overall conceptualization and interpretation of results, and contributed substantially to the writing of the manuscript. F.R contributed to data validation, interpretation of results, and the writing of the manuscript. W.J, J.L, and W.S contributed to discussions of the data and analysis and interpretation of results. All authors participated in the writing of the manuscript.

**Competing interests**

The authors declare that they have no competing interests.

**Correspondence** and requests for materials should be addressed to Hao Ding

**Supplementary Materials for “Satellite gravity constraints on inner-core viscosity and LLVPs density anomalies”**

Yachong An, Hao Ding*, Fred D. Richards, Weiping Jiang, Jiancheng Li, Wenbin Shen

*Email: dhaosgg@whu.edu.cn

**This PDF file includes:**

Supplementary text

Supplementary Figures S1 to S10

Supplementary Materials References

## Supporting Information Text

### 1 Definition of inertia moment and triaxiality parameter of the Earth's inner core

Considering both the flattening and degree-2 order-2 shape of the Earth, constant density surfaces of Earth are written as (Buffett, 1996)

$$r = a\left(1+\epsilon_{20}Y_{20}+\epsilon_{22}Y_{22}+\epsilon_{22}^{*}Y_{2-2}\right) \tag{S1}$$

where $\epsilon_{lm} = \epsilon_{lm}(a)$ denotes degree-$l$ order-$m$ shape amplitude varying with mean radius $a$. In particular, the relationship between the degree-2 order-0 amplitude and flattening is $\epsilon_{20} = -(16\pi/45)^{1/2}\bar{\epsilon}$ (or also written $f$ in geodesy). Where the asterisk ‘*’ represents complex conjugation; $Y_{lm} = Y_{lm}(\theta, \varphi)$ is degree-2 order-2 fully normalized spherical harmonic function. The axial moment of inertia, the mean equatorial moment of inertia, and the difference between the equatorial moments of inertia of the Earth are given by (Rochester et al., 2014)

$$C_{ic} = \frac{8\pi}{3}\int_0^{a_{ic}} \rho \left[ a^4 + \frac{2}{15}\frac{\mathrm{d}(a^5|\bar{\epsilon}|)}{\mathrm{d}a} \right] \mathrm{d}a \tag{S2a}$$

$$(A_{ic} + B_{ic})/2 = \frac{8\pi}{3}\int_0^{a_{ic}} \rho \left[ a^4 - \frac{1}{15}\frac{\mathrm{d}(a^5|\bar{\epsilon}|)}{\mathrm{d}a} \right] \mathrm{d}a \tag{S2b}$$

$$B_{ic} - A_{ic} = \sqrt{\frac{32\pi}{15}}\int_0^{a_{ic}} \rho \frac{\mathrm{d}(a^5|\epsilon_{22}|)}{\mathrm{d}a} \mathrm{d}a \tag{S2c}$$

respectively. The triaxiality parameter of the inner core is also used here, and it is defined as

$$\xi_s = \frac{B_s - A_s}{C_s - \dfrac{A_s + B_s}{2}} = \sqrt{\frac{15}{2\pi}}\int_0^{a_s} \rho \frac{\mathrm{d}(a^5|\epsilon_{2,2}|)}{\mathrm{d}a} \mathrm{d}a \Big/ \int_0^{a_s} \rho \frac{\mathrm{d}(a^5|\bar{\epsilon}|)}{\mathrm{d}a} \mathrm{d}a \tag{S3}$$

## 2 Evaluation of topographic and electromagnetic coupling at the CMB

The torque $-f_m(s)$ on the ends of the cylinders exerted by the mantle is further expressed as (Buffett & Mound, 2005)

$$f_m = \mathcal{F}_m(s)[\tilde{u}_f(s) - \tilde{u}_s] \tag{S4}$$

The amplitude $\mathcal{F}_m$ denote the coupling constant depended on the mechanism at the CMB. The electromagnetic coupling constant is written as (Dumberry & Mound, 2008)

$$\mathcal{F}_m^{EC}(s) = 2\pi s^3 \left[ (\hat{B}^r_{\mathrm{CMB}})^2 + (\bar{B}^r_{\mathrm{CMB}})^2 \right] G_m\left(\frac{a_f}{z_f}\right) \tag{S5}$$

In order to evaluate the CMB topographic coupling constant, in cylindrical coordinates, the cylindrical height of the CMB is expressed as (Buffett et al., 1998)

$$z_{\pm}(s,\phi) = \pm\left[z_f(s) - h_{\pm}(s,\phi)\right] \tag{S6}$$

Only antisymmetric structures about the equator are valid for CMB topographical coupling (Buffett et al., 1998), i.e.,

$$\begin{aligned} h_+ = -h_- = h^a(s,\phi) = h_c(s)\cos(m\phi) + h_s\sin(m\phi) \\ = h(s)\cos(m\phi + \phi_0) \end{aligned} \tag{S7}$$

When boundary topographical undulations are expressed as spherical harmonic expansions

$$h(\theta,\varphi) = \sum_{l,m} h_{l,m} Y_{l,m}(\theta,\varphi) \tag{S8}$$

When the sum of degree and order $l+m$ is odd and degree $m \neq 0$, it will be effective for terrain coupling. Topographic coupling constants are expressed as (Buffett et al., 1998)

$$\mathcal{F}_m^{TC}(s) = -2\pi m^2 h(s)^2 z_f \left[ i\omega s\rho - \frac{im^2}{s\omega}\frac{\{\bar{B}_\varphi^2\}}{\mu_0} \right] \tag{S9}$$

where $\{\bar{B}_\varphi^2\}$ represents the mean square value of the azimuthal component of the background magnetic field on a cylinder, which is set to $2\{B_s^2\}$ (Buffett et al., 1998).

Here, we use several widely disseminated models of CMB topography (Morelli & Dziewonski, 1987; Doornbos & Hilton, 1989; Steinberger & Holme, 2008; Yoshida, 2008; Soldati et al., 2013), as shown in Figure S1 (the corresponding degree-2 order-2 relief is also calculated and labeled below the model for subsequent explanation). The corresponding terrain coupling constant curve is shown in the Figure S3a. The largest one is from Morelli model (Morelli & Dziewonski, 1987), which reaches the order of $10^{17}$ N·m. For comparison, the electromagnetic coupling constant curve at the CMB is also plotted. The electromagnetic coupling constant reaches the order of $10^{23}$ N·m near the equator, which is ~6 orders larger than the topographic coupling constant, so the CMB topographic coupling effect can be ignored.

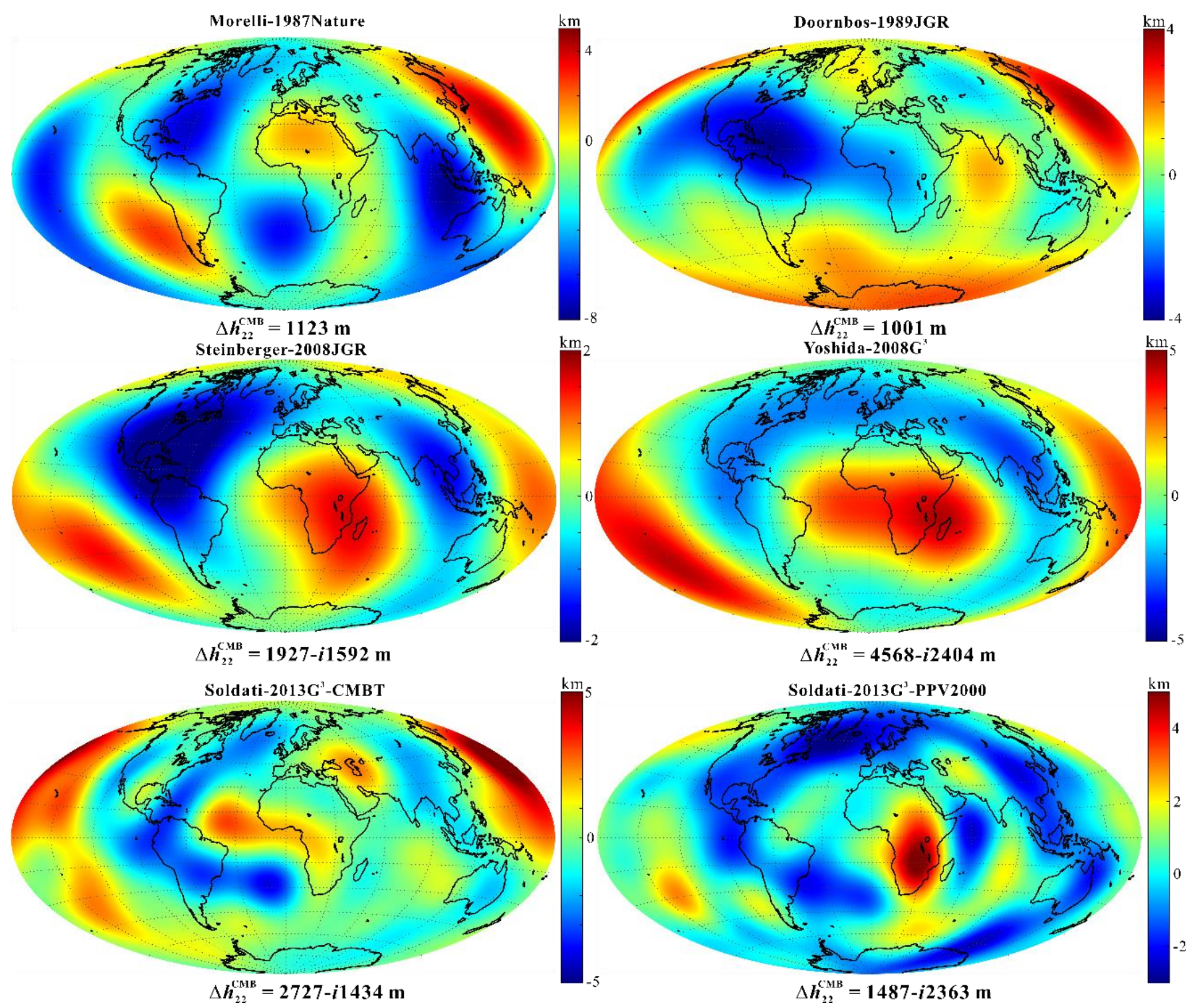

**Figure S1.** Models of the CMB topography derived from Morelli-1987Nature (Morelli & Dziewonski, 1987), Doornbos-1989JGR (Doornbos & Hilton, 1989), Steinberger-2008JGR (Steinberger & Holme, 2008), Yoshida-2008G$^3$ (Yoshida, 2008), Soldati-2013G$^3$-CMBT, and Soldati-2013G$^3$-PPV2000 (Soldati et al., 2013). We also calculate the height of the degree-2 order-2 relief in geographical coordinates, expressed in complex numbers.

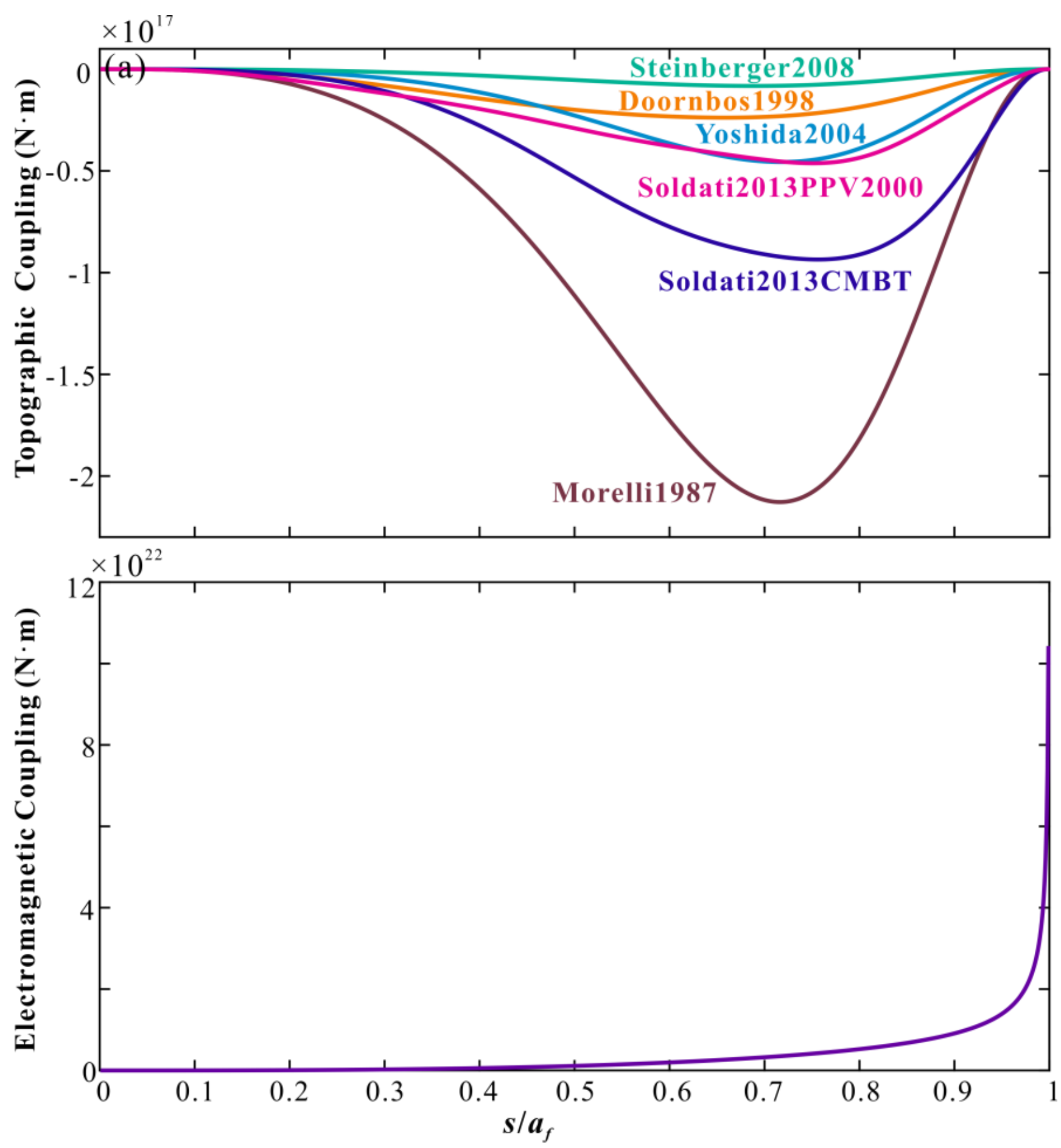

**Figure S2.** (a)Topographic coupling constants calculated based on the coefficients of the previous 6 core-mantle boundary topographic models as a function of radial distance *s* (imaginary part); (b) Electromagnetic coupling constants as a function of radial distance *s* (real part), where $G_m$ is set to be $1\times10^8$ S·m at the lower boundary.

## 3 Definition of the degree-2 order-2 interior type multipole belonging to mantle

The degree-2 order-2 multipole of the interior type belonging to the mantle is defined as (Chao, 2017)

$$q_{22}^{m} = \iiint_{\text{Mantle}} \rho(\mathbf{r}) r^{-1} Y_{2,2}(\Omega) \mathrm{d}r \mathrm{d}\Omega \tag{S10}$$

Considering the density heterogeneity of mantle, especially studying the related degree-2 order-2 lateral variation, 3-D density is written as

$$\rho(r,\theta,\varphi) = \rho(r)\left(1+\varepsilon_{2,2}Y_{2,2}+\varepsilon_{2,2}^{*}Y_{2,-2}\right) \tag{S11}$$

where $\varepsilon_{2,2}$ denotes the density anomaly coefficient of the spherical harmonic $Y_{2,2}(\theta, \varphi)$. Thus, the degree-2 order-2 interior type multipole belonging to mantle can be re-written as

$$q_{22}^{m} = \int_{a_f}^{a_e} \frac{\rho(a)}{a} \varepsilon_{2,2}^{*}(a)\mathrm{d}a \equiv \int_{a_f}^{a_e} p(a)\varepsilon_{2,2}^{*}(a)\mathrm{d}a \tag{S12}$$

According to Eq. (S12), the degree-2 order-2 density multipole $q_{22}^{m}$ can be regarded as the weighted integral of the amplitude of the degree-2 order-2 density anomaly $\varepsilon_{2,2}^{*}$ in the radial direction, and the normalized weight function $p(a)$ (gray curve) is shown in Figure S3, decreasing with the increase of radius $a$, which indicates that the density anomaly far away from the CMB has less and less influence on the multipole $q_{22}^{m}$, while the degree-2 order-2 density anomaly near the CMB corresponds to LLVPs found in seismic wave observation, so $q_{22}^{m}$ is sensitive to density anomaly of LLVPs.

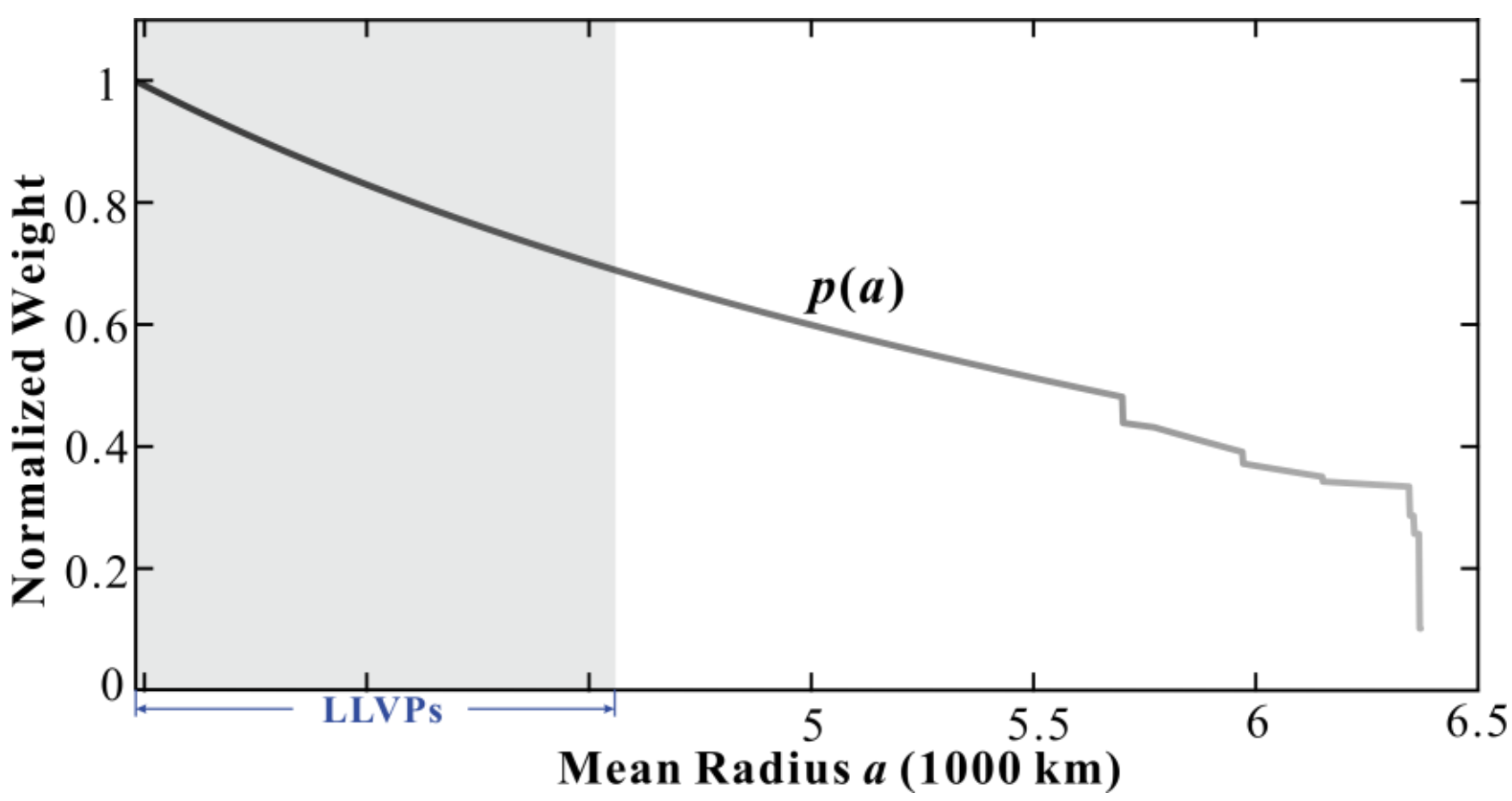

**Figure S3. Normalized weight function *p*(*a*)** (the gray shaded area indicates the LLVPs location range).

## 4 Shape of the Earth's core in hydrostatic equilibrium state

Although dense LLVPs can induce a local downward deflection of the CMB through internal loading, this effect represents only a short-wavelength and instantaneous mechanical response relative to a prescribed reference surface. A simple estimate based on internal-loading theory suggests that the associated degree-2 order-2 CMB depression is only on the order of $10^2$ m (Hager & O'Connell, 1981; Hager 1984; Richards & Hager, 1984), typically about –100 to –500 m depending on the adopted viscosity structure (Forte & Mitrovica, 1996; Peltier, 1999; Steinberger & Calderwood, 2006; Peltier et al., 2015; Lau et al., 2016; Roy & Peltier 2017). By contrast, over geological timescales, the CMB, which bounds a low-viscosity fluid outer core, tends to relax toward the hydrostatic and equipotential configuration of a self-gravitating Earth. In this equilibrium state, long-wavelength CMB topography is primarily controlled by global gravitational and rotational balance rather than by local mantle loading.

Similar to the Earth's flattening, the degree-2, order-2 shape $\epsilon_{22}(a)$ of the Earth's core is assumed to deviate little from hydrostatic equilibrium and should satisfies (Chao & Shih, 2024)

$$\frac{\mathrm{d}^2\epsilon_{22}}{\mathrm{d}a^2}+\frac{6}{a}\frac{\rho}{\bar{\rho}}\frac{\mathrm{d}\epsilon_{22}}{\mathrm{d}a}-\frac{6}{a^2}(1-\frac{\rho}{\bar{\rho}})\epsilon_{22}\approx 0 \tag{S13}$$

where the average density is defined as

$$\bar{\rho}(a)=\frac{3}{a^3}\int_0^a \rho(x)x^2\mathrm{d}x \tag{S14}$$

The profile of $\epsilon_{22}$ is shown in Figure S4, though this is not the true shape and requires scaling. Therefore, determining the scaling factor depends critically on the ratio of curvatures at the ICB and CMB, which can ultimately yield the true shape.

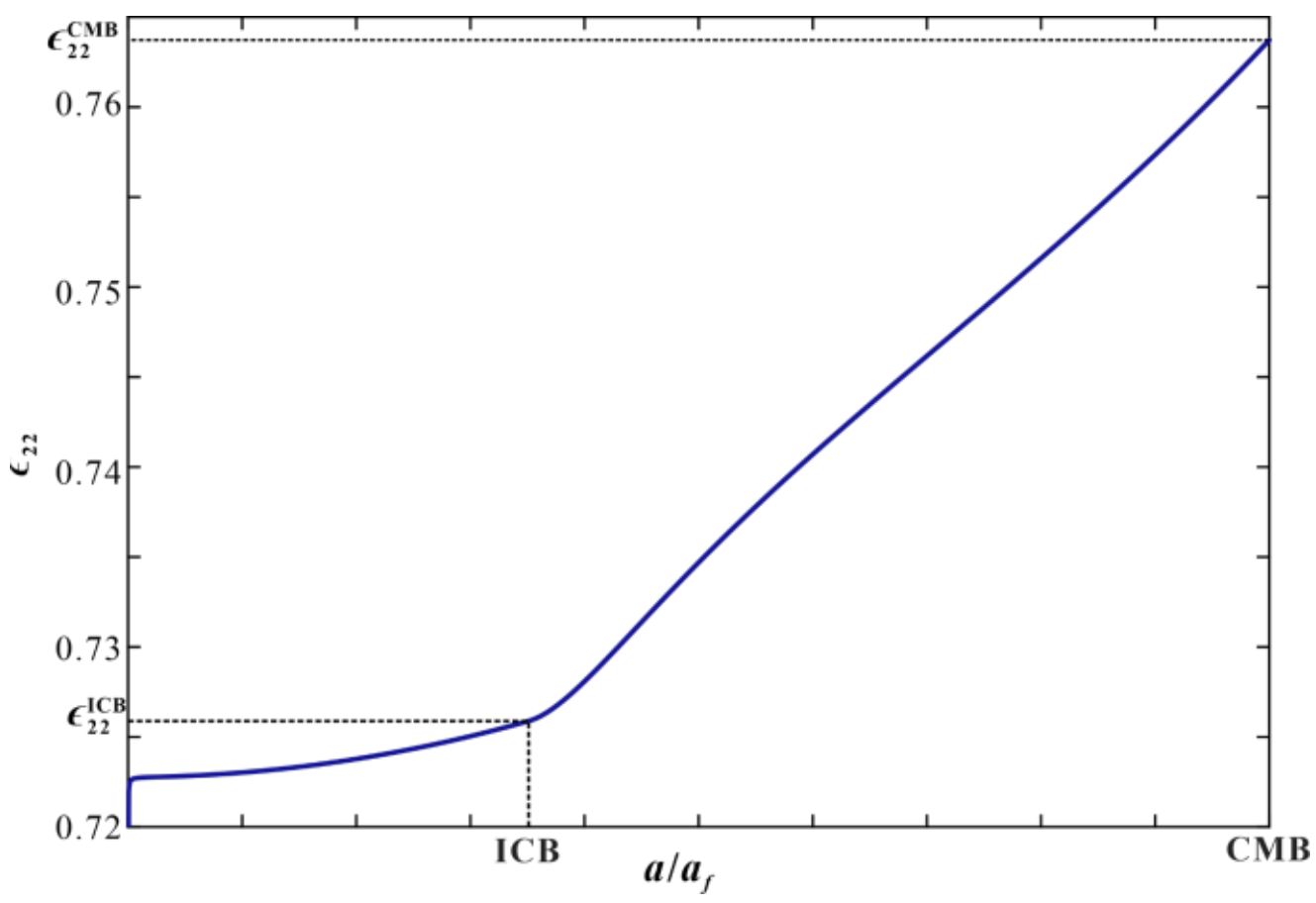

**Fig. S4. Radial profile of degree-2, order-2 shape in the core under hydrostatic equilibrium conditions** (Scaling required to obtain true shape.)

The degree-2, order-2 shape in Earth's core is mainly induced by the mantle (Chao & Shih, 2024),

$$q_{22}^{m} \approx \frac{5}{2}\sqrt{\frac{1}{\pi}}\epsilon_{22}^{\mathrm{CMB}}\frac{Q_{00}^{\mathrm{CMB}}}{a_f^3} - \frac{Q_{22}^{\mathrm{CMB}}}{a_f^5} \quad \text{(S15)}$$

with

$$Q_{00}(a) = \sqrt{4\pi}\int_0^a \rho(x)x^2\mathrm{d}x \quad \text{(S16)}$$

$$Q_{22}(a) = \int_0^a \rho(x)\frac{\mathrm{d}(x^5\epsilon_{22})}{\mathrm{d}x}\mathrm{d}x = \sqrt{\frac{15}{32\pi}}[B(a) - A(a)] \quad \text{(S17)}$$

Here, $Q_{00}(a)$ depends only on Earth's density structure and equals $Q_{00}^{\mathrm{CMB}}=5.4714\times10^{23}$ at the CMB, while $Q_{22}(a)$ depends on the shape profile $\epsilon_{22}(a)$. Since the scaling factor is unknown, it cancels out in the ratio, allowing one to derive $Q_{22}^{\mathrm{CMB}} / \epsilon_{22}^{\mathrm{CMB}} = 5.4419\times10^{36}$. From the shape profile in Figure S4 and PREM model, we can get

$$q_{22}^{m} = 8.048\times10^3\epsilon_{22}^{\mathrm{ICB}} \equiv o_2\epsilon_{22}^{\mathrm{ICB}} \quad \text{(S18)}$$

Similarly, from the shape profile, we obtain $Q_{22}^{ic} / \epsilon_{22}^{\mathrm{ICB}} = 3.496 \times 10^{34}$, that is

$$Q_{22}^{ic} = 3.496 \times 10^{34} \epsilon_{22}^{\mathrm{ICB}} \equiv o_3 \epsilon_{22}^{\mathrm{ICB}} = \frac{1}{8}\sqrt{\frac{15}{2\pi}} \xi_{ic} (A_{ic} + B_{ic}) e_{ic} \tag{S19}$$

And finally

$$q_{22}^{m} = \frac{o_2}{o_3} Q_{22}^{ic} = \frac{1}{8}\sqrt{\frac{15}{2\pi}} (A_{ic} + B_{ic}) e_{ic} \frac{o_2}{o_3} \xi_{ic} \equiv o_1 \xi_{ic} \tag{S20}$$

where $o_1 = \frac{1}{8}\sqrt{\frac{15}{2\pi}} (A_{ic} + B_{ic}) e_{ic} \frac{o_2}{o_3} = 12.606 \ \mathrm{kg} \cdot \mathrm{m}^{-3}$.

Under hydrostatic equilibrium, we estimate that mantle density anomalies can produce a CMB topographic relief of up to ~+800 m in the main text, and the predicted outward and positive long-wavelength CMB relief is also close to Morelli-1987Nature (Morelli & Dziewonski, 1987) and Doornbos-1989JGR (Doornbos & Hilton, 1989) seismic CMB topography models (Figure S1).

## 5 Length-of-day variation observation and external excitation sources

The length-of-day variation (ΔLOD) observation spanning from 1962/1 to 2022/6, with daily sampling, are depicted in Figure S5, alongside the corresponding results excited by atmospheric angular momentum (AAM; spanning from 1948/1 to 2021/1, with 6-hour sampling), oceanic angular momentum (OAM; spanning from 1949/1 to 2020/2, with 10-day sampling), and hydrological angular momentum (HAM; spanning from 1971/1 to 2022/7, with daily sampling). Their respective Fourier amplitude spectra are presented in Figure S5b. Notably, within periods shorter than 5 years, the amplitude of ΔLOD observation and this excited by AAM and OAM exhibit remarkable similarity (illustrated by the black fine curve and gray thick curve, respectively).

Subsequent removal of atmospheric and oceanic effects (denoted by the red curve) results in the elimination of nearly all signals, with residual noise closely resembling hydrological effects (as depicted by the green curve). Thus, signals with periods shorter than 5 years predominantly originate from external sources, and the candidate signal for the inner core oscillation is not in this band. Interestingly, upon removal of atmospheric and oceanic effects, signals on the intradecadal scale exhibit amplification (illustrated by the gray thick curve and red curve), while the contribution of hydrological effect remains significantly lower than background noise (as indicated by the green curve). Consequently, signals on the intradecadal scale primarily arise from the Earth's interior, representing a suspicious frequency band for investigating inner core oscillation. On a decade-scale or longer, although signals are attenuated following atmospheric and oceanic removal, background noise surpasses hydrological effect. Therefore, the possibility of inner core oscillation cannot be dismissed within this frequency band, prompting further investigation to exclude inner core oscillation through combined analysis of gravitational field observations and theoretical model.

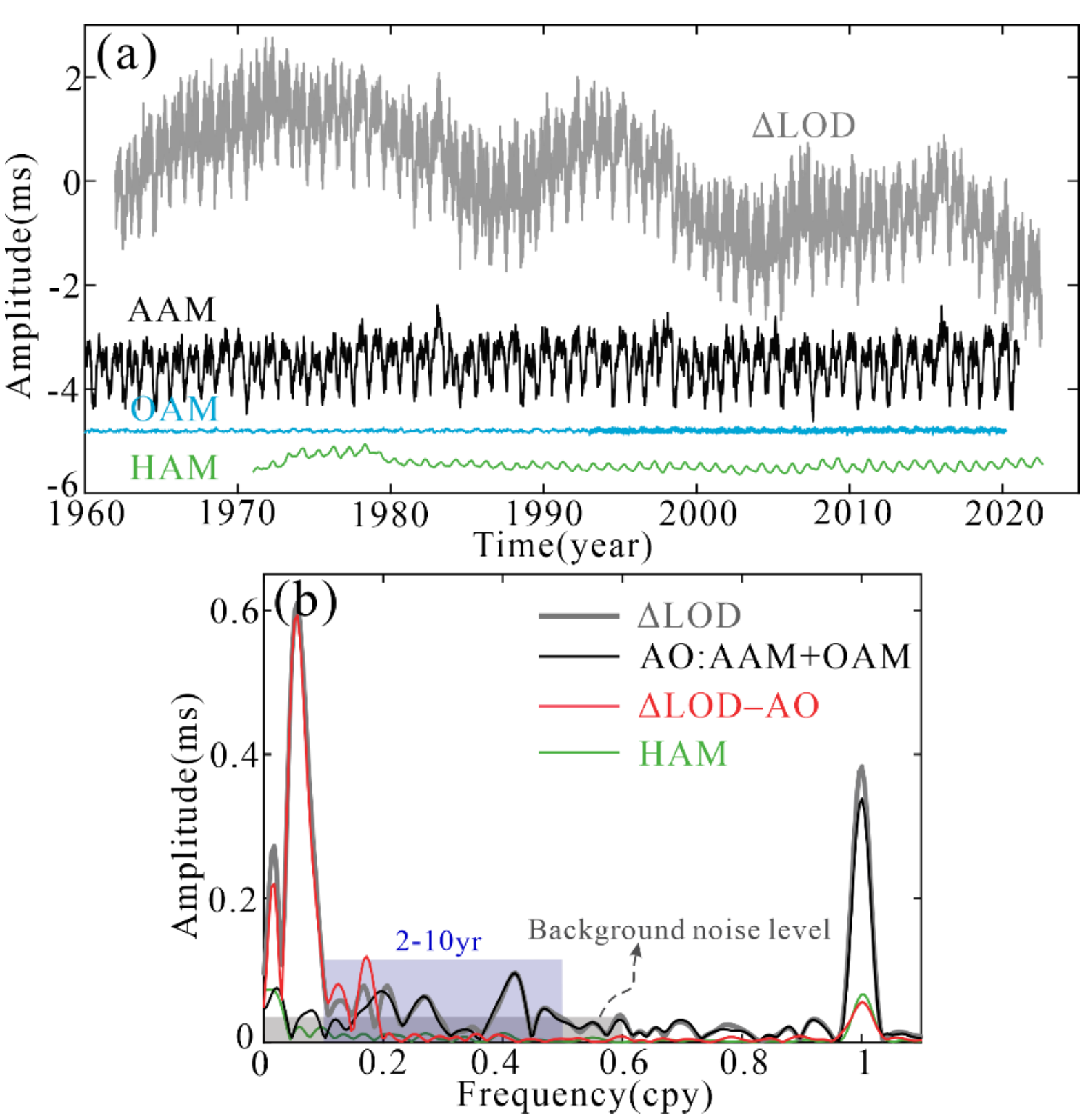

**Figure S5. Length-of-day variation (ΔLOD) from observation and external source excitations and corresponding Fourier spectra**. (a) Length-of-day variation (ΔLOD) observation with their corresponding excited results by the atmospheric, oceanic (OAM), and hydrologic angular momentum (AAM, OAM, and HAM), respectively. (b) Fourier amplitude spectra of the ΔLOD observation, corresponding excited results by the AAM and OAM (AAM+OAM; AO) and HAM, and the residual sequence of ΔLOD observation minus AAM and OAM (ΔLOD−AO).

## 6 Verify detection results of the degree-2 order-2 Stokes coefficients

To ensure the robustness of the $\Delta C_{22}$ and $\Delta S_{22}$ signals, in addition to the CSR_SLR dataset used in the main text, we also analyze the IGG_SLR dataset (Löcher & Kusche, 2020) and employ two independent hydrological models, GLDAS and ISBA (Decharme et al., 2019), for correction. Figure S6 shows the results from CSR_SLR corrected using ISBA (panel a), and from IGG_SLR corrected using ISBA (panel b) and GLDAS (panel c). The results are highly consistent across datasets and correction schemes. In all cases, a clear ~6-year signal is observed in $\Delta S_{22}$. In contrast, no significant ~5-6-year signal is detected in $\Delta C_{22}$ for Figures S6a and S6b, while Figure S6c shows only a relatively weak peak near ~5.3 years. A similar ~5.3-year feature is also present in the GLDAS-corrected CSR_SLR result shown in the main text, but has already been demonstrated to be non-stationary (Figure 3b).

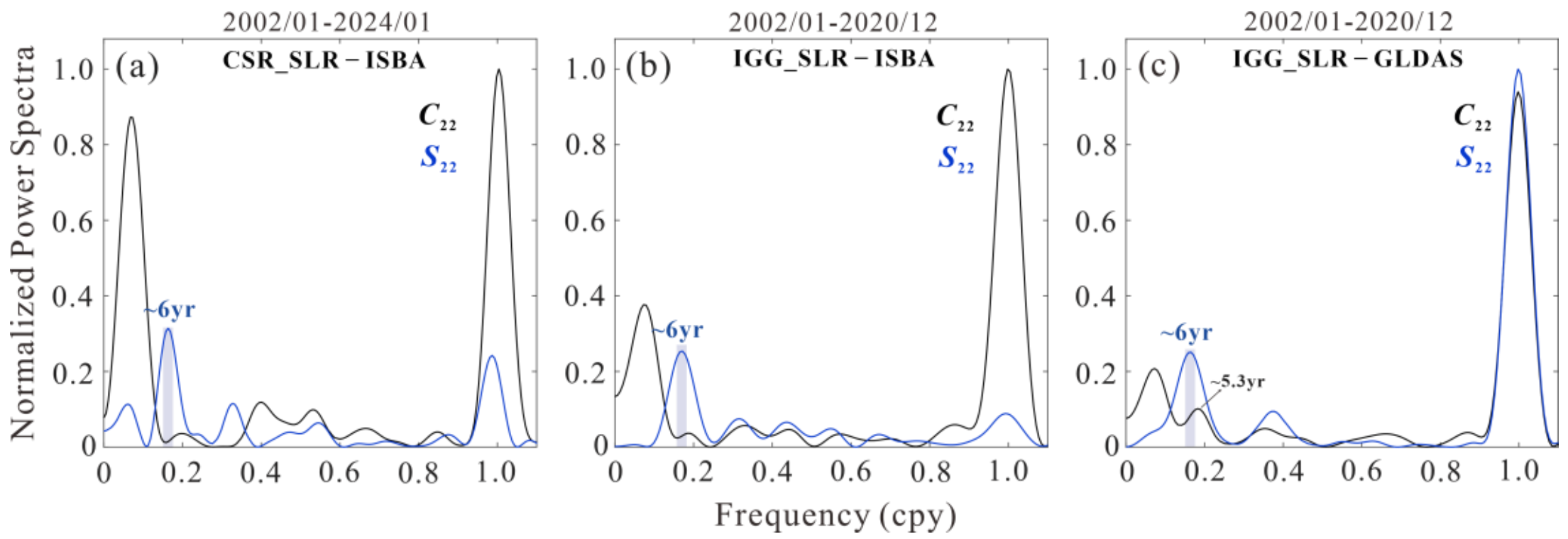

**Figure S6. The Fourier power spectra of products** removed different hydrological models for **the $\Delta C_{22}/\Delta S_{22}$.** (a) CSR_SLR after removing the ISBA hydrological model. (b) IGG_SLR after removing the ISBA hydrological model. (c) IGG_SLR after removing the GLDAS hydrological model.

Figure S7 also presents the wavelet spectra of the CSR_SLR results corrected using the two hydrological models. A persistent ~6-year signal is clearly visible in $\Delta S_{22}$, whereas no stable ~6-year signal is found in $\Delta C_{22}$. Instead, the GLDAS-corrected result exhibits a relatively strong, non-stationary oscillation near 4.5-year. This transient signal likely accounts for the apparent ~5.3-year spectral peak observed in Figure 3a of the main text and in Figure S7c. Because the Fourier spectrum represents an average over time, the superposition of this non-stationary ~4.5-year variability with a very weak ~6-year component in $\Delta C_{22}$ produces an apparent peak near ~5.3-year.

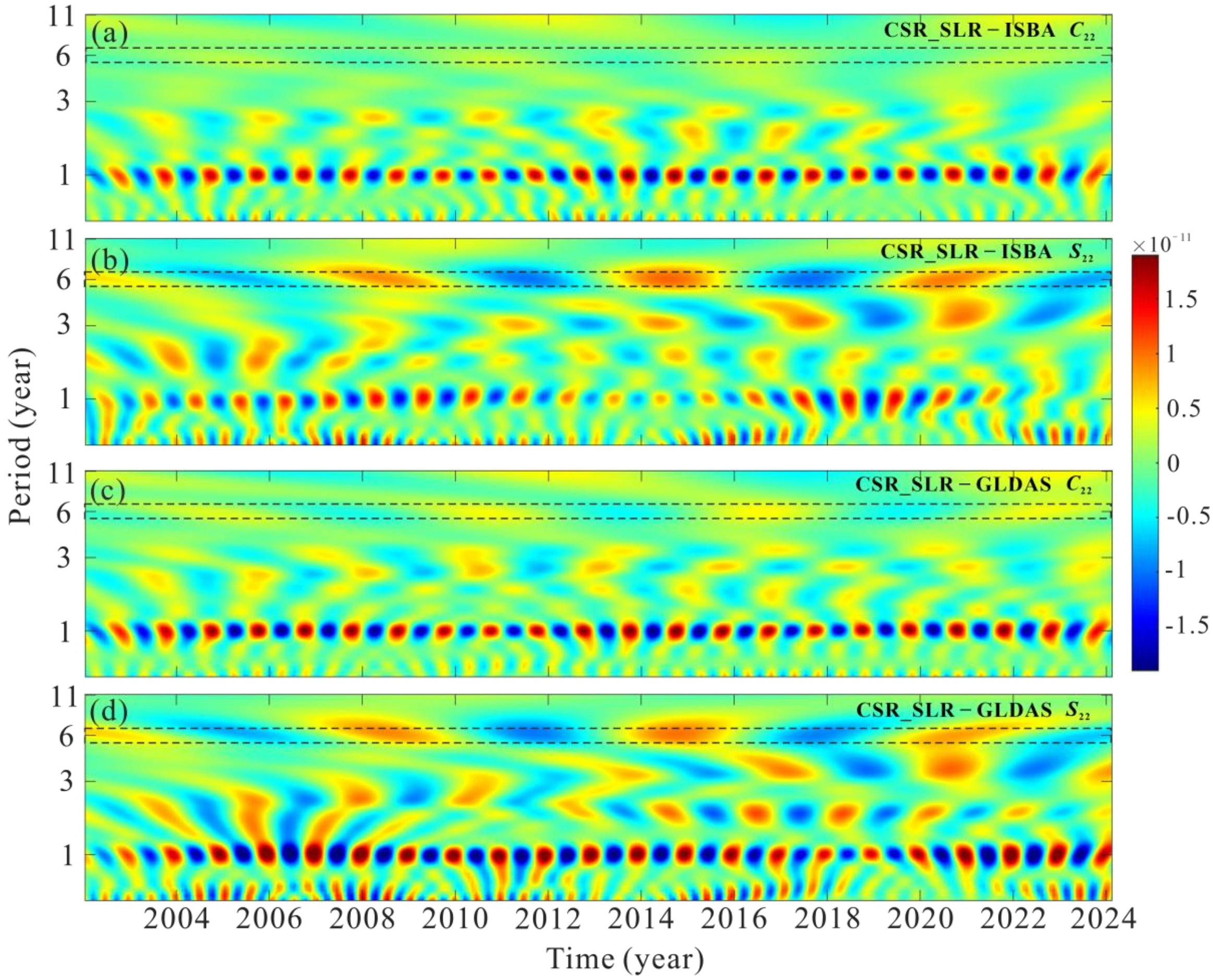

**Figure S7. Wavelet spectra of the degree-2 order-2 gravitational Stokes coefficients that the CSR_SLR product remove ISBA hydrological model for** (a) the $\Delta C_{22}$ (b) the $\Delta S_{22}$ and **CLDAS hydrological model for** (c) the $\Delta C_{22}$ (d) the $\Delta S_{22}$.

The Figure S8 shows the contributions of the ISBA and GLDAS hydrological models to $\Delta C_{22}$ and $\Delta S_{22}$. No ~6-year signal is observed in either case. However, the variability in the 2-11 years band is noticeably stronger in the ISBA-based correction than in the GLDAS-based result. This difference also explains why the GLDAS-corrected $\Delta C_{22}$ results shown in the previous figures exhibit relatively higher residual spectral peaks.

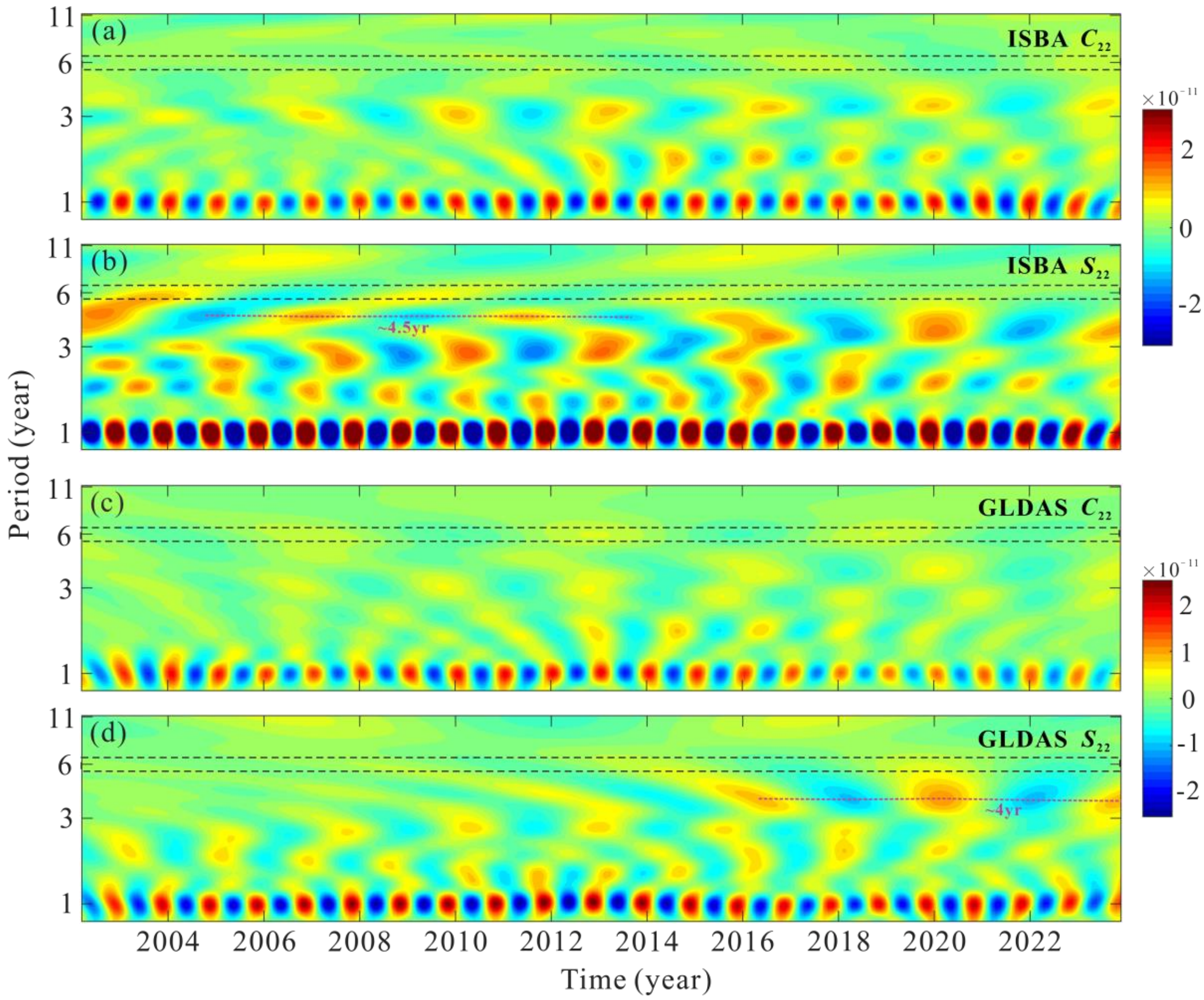

**Figure S8. Wavelet spectra of the degree-2 order-2 gravitational Stokes coefficients of ISBS hydrological model for** (a) the $\Delta C_{22}$ (b) the $\Delta S_{22}$ and **CLDAS hydrological model for** (c) the $\Delta C_{22}$ (d) the $\Delta S_{22}$.

Overall, these three figures demonstrate that no ~6-year oscillation is present in the hydrological contributions. Therefore, the ~6-year signal identified in $\Delta S_{22}$ is robust. The weighted mean amplitude is estimated to be ~$1.08\times10^{-11}$.

**7 Numerical simulation case**

In this case, the parameters set are: the RMS of axial dipole component and nondipole component of radial magnetic field across the CMB are set to 0.226 mT and 0.42 mT, respectively; they are set to 2 mT and 3 mT across the ICB, respectively; the conductivity of the fluid core is set to $\sigma_f$ = 5×10$^5$ S·m$^{-1}$; the total conductance of the mantle is $G_m$ = 3×10$^8$ S·m$^{-1}$; the relaxation time of core viscosity is set to $t_\tau$ = 0.682 years.

Figure S9a shows the ratios $\zeta$ in the eigen-function for the 5.9-year eigen-mode solved when the MICG constant $\hat{\Gamma}_z$ is adjusted (pink curve), where $\hat{\Gamma}_z$ is set in the range 1.6-2×10$^{20}$ N·m; the relationship from observed ΔLOD and $\Delta S_{22}$ data is also shown Figure S9a (pink curve). Figure S9b shows the ratio of the real and imaginary parts of the ratio $\zeta$ solved. When the MICG constant $\hat{\Gamma}_z$ = 1.8×10$^{20}$ N·m, the solution curve intersects with the observation curve exactly, and the real part is much larger than the imaginary part. So from observation and simulation we match to $t_\tau$ = 0.682 years and t $\hat{\Gamma}_z$ = 1.8×10$^{20}$ N·m.

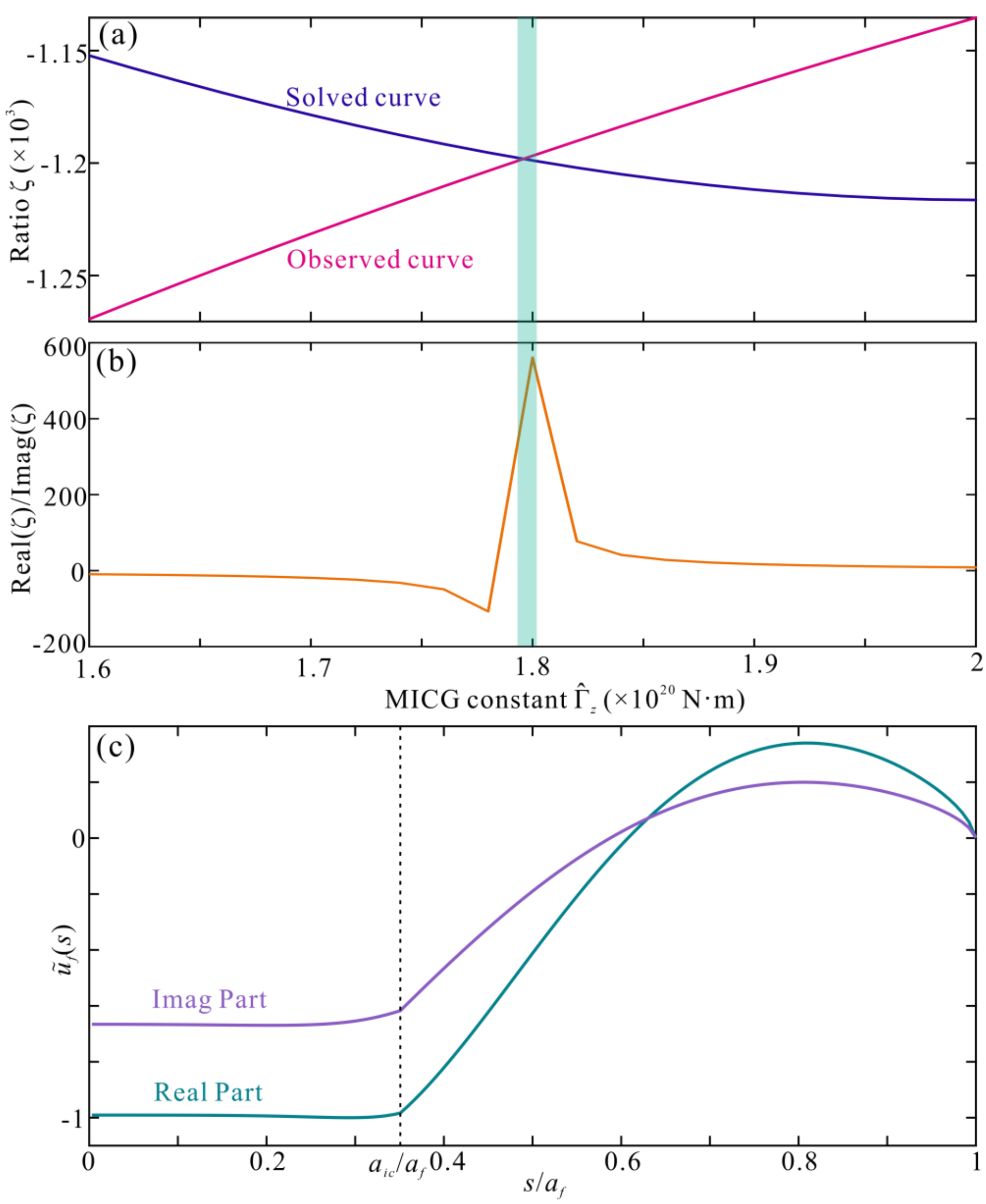

**Figure S9. Example of numerical solution for torsion system in earth core.** The real part (a) and the ratio of real part to imaginary part (b) of the ratio $\zeta$ from the solved 5.9-year eigen-function vary with MICG constant $\hat{\Gamma}_z$ when the relaxation time of core viscosity $t_\tau$ is fixed at 0.682 years. (c) The real and imaginary parts of the solved normalized angular velocity eigen-function $\tilde{u}_f$ of the fluid core as a function of radial distance $s$ in cylindrical coordinate; the angular velocity amplitude of the solved inner core $\tilde{u}_s$ is also shown in the figure, which coincides with $\tilde{u}_f(a_s)$ at the equator of ICB (since the electromagnetic coupling constant tends to infinity there), verifying the correctness of the results, the viscous relaxation time of the inner core $t_\tau$ and MICG constant $\hat{\Gamma}_z$ are set to 0.682 years and $1.8 \times 10^{20}$ N·m, respectively.

### 8 Density anomaly inversion based on S362MANI model

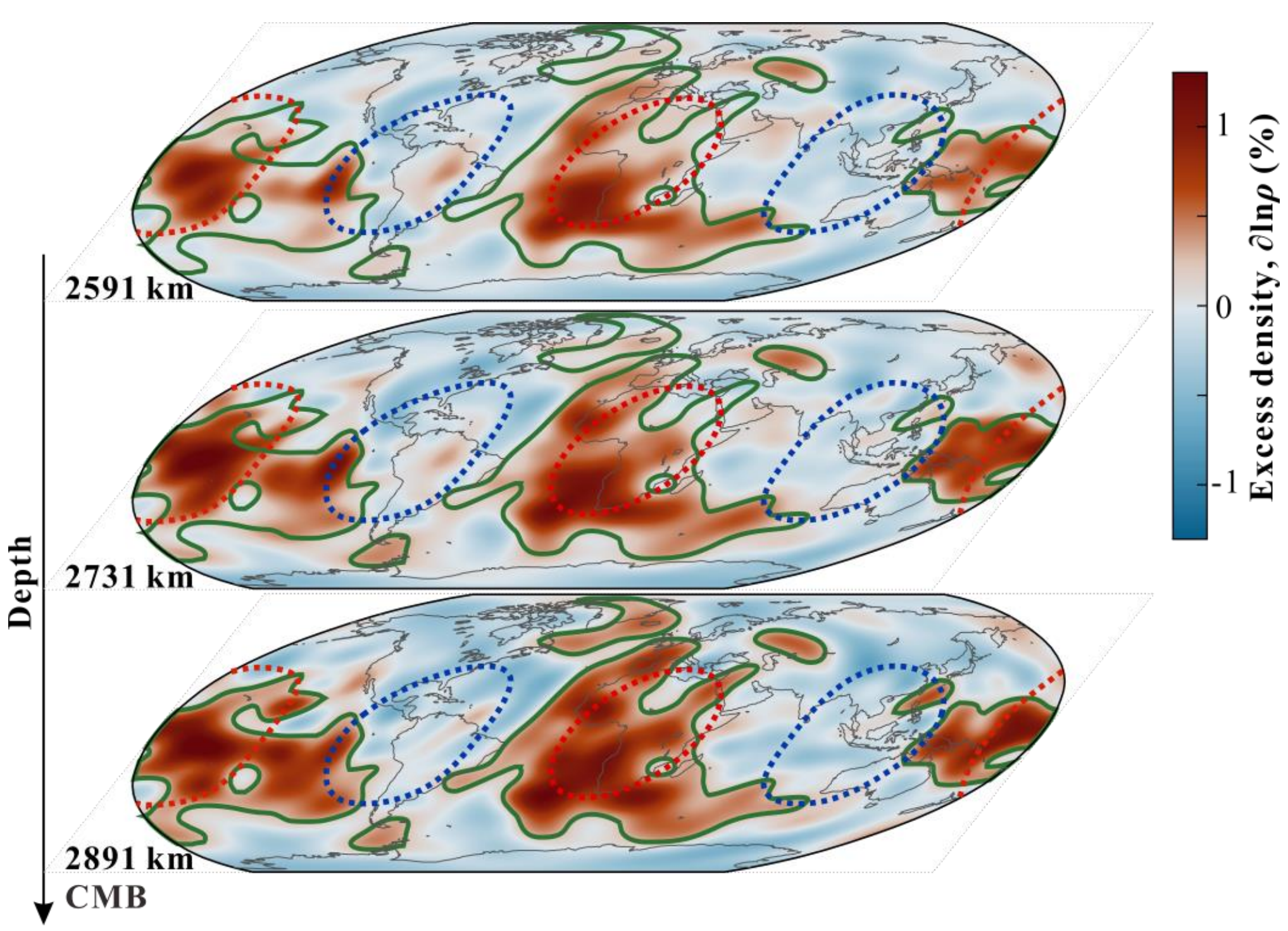

**Figure S10. Inversed density distribution map at depth 2891 km, 2731 km, and 2591km based on S362MANI model (Kustowski et al., 2008).**

### Supplementary Materials References